\documentclass[11pt]{article}
\usepackage[font=small,labelfont=bf]{caption}
\usepackage{subcaption}
\usepackage{amsmath,amscd,amssymb,mathrsfs,bm}
\usepackage{mathtools}
\usepackage{cancel}
\DeclareSymbolFontAlphabet{\mathrsfs}{rsfs}
\usepackage[mathscr]{eucal}
\usepackage{mathabx}
\usepackage{tikz}
\usepackage{tikz-cd}
\usepackage{tikz-feynman}
\usetikzlibrary{decorations.markings}
\usepackage[normalem]{ulem} 
\usetikzlibrary{quotes}
\usepackage{savesym}
\usepackage[nosort]{cite}
\usepackage{wasysym}
\usepackage{graphicx}
\usepackage{pifont}
\usepackage{float}
\usepackage{multicol}
\usepackage{amsfonts}
\usepackage{picture}

\savesymbol{iint}
\savesymbol{iiint}
\restoresymbol{TXF}{iint}
\restoresymbol{TXF}{iiint}
\usepackage{xcolor}
\usepackage{setspace}
\usepackage{clay}
\usepackage{slashed}
\usepackage{hyperref}
\hypersetup{colorlinks=true}
\hypersetup{linkcolor=black}
\hypersetup{citecolor=magenta}
\hypersetup{urlcolor=blue}
\usepackage{multirow}
\usepackage{verbatim}
\usepackage{accents}
\usepackage{placeins}
\usepackage{enumitem}
\usepackage[all]{xy}
\definecolor{refkey}{rgb}{0,.8,.2}\definecolor{labelkey}{rgb}{1,0,0}
\numberwithin{equation}{section}

\graphicspath{ {./Figs/} }

\newenvironment{nalign}{
	\begin{equation}
	\begin{aligned}
}{
	\end{aligned}
	\end{equation}
	\ignorespacesafterend
}

\usepackage{changepage}                 

\newlength{\offsetpage}
{\end{adjustwidth}}
\newlength{\offsetpageII}
\newenvironment{widepageII}{\begin{adjustwidth}{-\offsetpageII}{-\offsetpageII}%
    \addtolength{\textwidth}{2\offsetpageII}}%
{\end{adjustwidth}}

\newlength{\offsetpageIII}
\newenvironment{widepageIII}{\begin{adjustwidth}{-\offsetpageIII}{-\offsetpageIII}%
    \addtolength{\textwidth}{2\offsetpageIII}}%
{\end{adjustwidth}}

\usepackage{bm}

\def\betaR{\boldsymbol{\beta}_R^a}
\def\betaLR{\boldsymbol{\beta}_{LR}^a}
\def\betaRbetaR{\betaR\otimes\betaR}
\def\betaRR{\boldsymbol{\beta}_{RR}^a}
\def\Z{\mathbb{Z}}

\def\C{\mathbb{C}}

\def\g{\mathfrak{g}}

\def\GUV{G_{\mathrm{EWF}}}
\def\FSS{SU(4)\times Sp(6)_L \times Sp(6)_R}
\def\GSM{G_{\mathrm{SM}}}
\def\Gint{G_{\mathrm{int}}}
\def\GintE{SU(3)\times \prod_i SU(2)_{L,i} \times Sp(4)_{R,12}\times U(1)_R}
\def\SOTen{\mathrm{Spin}(10)}

\def\su{\mathfrak{su}}
\def\sp{\mathfrak{sp}}
\def\so{\mathfrak{so}}

\def\ZP{Z^\prime}
\def\Tr{\mathrm{Tr~}}

\def\CKM{V_{\mathrm{CKM}}}

\usepackage{xspace}

\newcommand{\be}{\begin{equation}}
\newcommand{\ee}{\end{equation}}
\newcommand{\bea}{\begin{eqnarray}}
\newcommand{\eea}{\end{eqnarray}}
\newcommand{\beq}{\begin{equation}}
\newcommand{\eeq}{\end{equation}}

\newcommand{\cL}{{\cal L}}

\newcommand{\cO}{{\cal O}}

\definecolor{purple}{RGB}{127, 0, 255}
\definecolor{orange}{RGB}{255,165,0}
\definecolor{red1}{RGB}{255,0,0}
\definecolor{red2}{RGB}{139,0,0}
\definecolor{orange1}{RGB}{255,165,0}
\definecolor{orange2}{RGB}{255,140,0}
\definecolor{yellow1}{RGB}{240,230,140}
\definecolor{yellow2}{RGB}{189,183,107}
\definecolor{green1}{RGB}{0,128,0}
\definecolor{green2}{RGB}{0,100,0}
\definecolor{blue1}{RGB}{0,0,255}
\definecolor{blue2}{RGB}{0,0,139}
\definecolor{indigo1}{RGB}{75,0,130}
\definecolor{indigo2}{RGB}{138,43,226}
\definecolor{viloet1}{RGB}{238,130,238}
\definecolor{viloet2}{RGB}{255,0,255}
\definecolor{turquise1}{RGB}{64,224,208}
\definecolor{turquise2}{RGB}{0,206,209}
\definecolor{brown1}{RGB}{205,133,63}
\definecolor{brown2}{RGB}{244,164,96}
\definecolor{gray1}{RGB}{128,128,128}
\definecolor{gray2}{RGB}{211,211,211}
\definecolor{gold1}{RGB}{255,215,0}
\definecolor{gold2}{RGB}{218,165,32}
\definecolor{black1}{RGB}{0,0,0}
\definecolor{black2}{RGB}{105,105,105}

\newcommand{\colordiagram}[1]{
\begin{subfigure}[b]{0.31\textwidth}
\begin{tikzpicture}
\begin{feynman}
#1
\end{feynman}
\end{tikzpicture}
\end{subfigure}
}
\def\fermions{
    \vertex [] (psi1) {\footnotesize $\overline{\Psi}_L$};
    \vertex [right=0.8in of psi1] (c0);
    \vertex [right=1.6in of psi1] (psi2) {\footnotesize $\Psi_R$};
    \vertex [above left =0.05in and 0.05in of c0.east] (L0);
    \vertex [above right =0.05in and 0.05in of c0.west] (R0);
    \vertex [above  =0.05in of psi1.east] (psi1V);
    \draw[red]  (L0)--(psi1V);
    \vertex [above  =0.05in of psi2.west] (psi2V);
    \draw[blue,dash pattern={on 3pt off 1pt}]  (psi2V)--(R0);
    \diagram* {
    (psi1) -- [fermion] (c0)  -- [anti fermion] (psi2)
    };
}
\def\sep{0.35in}

\newcommand{\HiggsII}[2]{
    \vertex [above=\sep of c0] (c1);
    \vertex [above=\sep of c1] (c2) {\footnotesize $\langle H_a\rangle$};
    \diagram* {(c0)--[scalar] (c1)--[scalar] (c2)};
    \vertex [ left  =0.05in of c2.south] (#1);
    \vertex [ right  =0.05in of c2.south] (#2);
}
\newcommand{\HiggsIII}[2]{
    \vertex [above=\sep of c0] (c1);
    \vertex [above=\sep of c1] (c2);
    \vertex [above=\sep of c2] (c3) {\footnotesize $\langle H_a\rangle$};
    \diagram* {(c0)--[scalar] (c1)--[scalar] (c2)--[scalar] (c3)};
    \vertex [ left  =0.05in of c3.south] (#1);
    \vertex [ right  =0.05in of c3.south] (#2);
}
\newcommand{\HiggsIV}[2]{
    \vertex [above=\sep of c0] (c1);
    \vertex [above=\sep of c1] (c2);
    \vertex [above=\sep of c2] (c3);
    \vertex [above=\sep of c3] (c4) {\footnotesize $\langle H_a\rangle$};
    \diagram* {(c0)--[scalar] (c1)--[scalar] (c2)--[scalar] (c3)--[scalar] (c4)};
    \vertex [ left  =0.05in of c4.south] (#1);
    \vertex [ right  =0.05in of c4.south] (#2);
}
\newcommand{\HiggsV}[2]{
    \vertex [above=\sep of c0] (c1);
    \vertex [above=\sep of c1] (c2);
    \vertex [above=\sep of c2] (c3);
    \vertex [above=\sep of c3] (c4);
    \vertex [above=\sep of c4] (c5) {\footnotesize $\langle H_a\rangle$};
    \diagram* {(c0)--[scalar] (c1)--[scalar] (c2)--[scalar] (c3)--[scalar] (c4)--[scalar] (c5)};
    \vertex [ left  =0.05in of c5.south] (#1);
    \vertex [ right  =0.05in of c5.south] (#2);
    }
\newcommand{\phiLI}[1]{
    \vertex [left=0 in of #1] (A);
    \vertex [below left =0.02in and 0.05in of A.west] (L1);
    \vertex [above left =0.02in and 0.05in of A.west] (L2);
    \vertex [left=0.3in of A] (E) {\footnotesize $\langle\Phi_L\rangle$};
    \diagram* {(A) --[scalar] (E)};
    \vertex [left=0.27in of L1] (B);
    \draw[red] (B)--(L1);
    \vertex [left=0.27in of L2] (D);
    \draw[red] (L2)-- (D);
}
\newcommand{\phiLII}[1]{
    \vertex [left=0 in of #1] (A);
    \vertex [below left =0.02in and 0.05in of A.west] (L3);
    \vertex [above left =0.02in and 0.05in of A.west] (L4);
    \vertex [left=0.3in of A] (E) {\footnotesize $\langle\Phi_L\rangle$};
    \diagram* {(A) --[scalar] (E)};
    \vertex [left=0.27in of L3] (B);
    \draw[red] (B)--(L3);
    \vertex [left=0.27in of L4] (D);
    \draw[red] (L4)-- (D);
}

\newcommand{\phiRI}[1]{
    \vertex [right=0 in of #1] (A);
    \vertex [below right =0.02in and 0.05in of A.east] (R1);
    \vertex [above right =0.02in and 0.05in of A.east] (R2);
    \vertex [right=0.3in of A] (E) {\footnotesize $\langle\Phi_R\rangle$};
    \diagram* {(A) --[scalar] (E)};
    \vertex [right=0.27in of R1] (B);
    \draw[blue,dash pattern={on 3pt off 1pt}] (B)--(R1);
    \vertex [right=0.27in of R2] (D);
    \draw[blue,dash pattern={on 3pt off 1pt}] (D)-- (R2);
}
\newcommand{\phiRII}[1]{
    \vertex [right=0 in of #1] (A);
    \vertex [below right =0.02in and 0.05in of A.east] (R3);
    \vertex [above right =0.02in and 0.05in of A.east] (R4);
    \vertex [right=0.3in of A] (E) {\footnotesize $\langle\Phi_R\rangle$};
    \diagram* {(A) --[scalar] (E)};
    \vertex [right=0.27in of R3] (B);
    \draw[blue,dash pattern={on 3pt off 1pt}] (B)--(R3);
    \vertex [right=0.27in of R4] (D);
    \draw[blue,dash pattern={on 3pt off 1pt}] (D)-- (R4);
}

\newcommand{\phiLL}[1]{
    \vertex [left=0 in of #1] (A);
    \vertex [above left=0.05in and 0.5in of A] (A2) {\footnotesize $\langle\Phi_L\rangle$};
    \vertex [below left=0.05in and 0.5in of A] (A1) {\footnotesize $\langle\Phi_L\rangle$};
    \diagram* {(A) --[scalar] (A1),(A) --[scalar] (A2)};
    \vertex [below left =0.04in and 0.05in of A.east] (L1);
    \vertex [above left =0.04in and 0.05in of A.east] (L2);
    \vertex[below left=0.11in and 0.45in of L1] (B1);
    \draw[red] (L1)-- (B1);
    \vertex[above left=0.11in and 0.45in of L2] (B2);
    \draw[red] (L2)-- (B2);
    \vertex[above=0.08in of B1] (D1);
    \vertex[below=0.08in of B2] (D2);
    \vertex[left=0.14in of A] (E);
    \draw[red] (D1)--(E);
    \draw[red,postaction={decoration={markings,mark=at position 0.7 with {\arrow{>>}}},decorate}] (D2)--(E);
}
\newcommand{\phiRR}[1]{
    \vertex [right=0 in of #1] (A);
    \vertex [above right=0.05in and 0.5in of A] (A2) {\footnotesize $\langle\Phi_R\rangle$};
    \vertex [below right=0.05in and 0.5in of A] (A1) {\footnotesize $\langle\Phi_R\rangle$};
    \diagram* {(A) --[scalar] (A1),(A) --[scalar] (A2)};
    \vertex [below right =0.04in and 0.05in of A.west] (R1);
    \vertex [above right =0.04in and 0.05in of A.west] (R2);
    \vertex[below right=0.11in and 0.45in of R1] (B1);
    \draw[dash pattern={on 3pt off 1pt},blue] (B1)-- (R1);
    \vertex[above right=0.11in and 0.45in of R2] (B2);
    \draw[blue,dash pattern={on 3pt off 1pt}] (B2)-- (R2);
    \vertex[above=0.08in of B1] (D1);
    \vertex[below=0.08in of B2] (D2);
    \vertex[right=0.14in of A] (E);
    \draw[blue,dash pattern={on 3pt off 1pt}] (D1)--(E);
    \draw[dash pattern={on 3pt off 1pt},blue,postaction={decoration={markings,mark=at position 0.7 with {\arrow{>>}}},decorate}] (D2)--(E);
}
\newcommand{\LIII}[2]{
\draw[red,postaction={decoration={markings,mark=at position 0.6 with {\arrow{<<<}}},decorate}]  (#1)-- (#2);
}
\newcommand{\LII}[2]{
\draw[red,postaction={decoration={markings,mark=at position 0.5 with {\arrow{<<}}},decorate}]  (#1)-- (#2);
}
\newcommand{\LI}[2]{
\draw[red,postaction={decoration={markings,mark=at position 0.5 with {\arrow{<}}},decorate}]  (#1)-- (#2);
}
\newcommand{\RIII}[2]{
\draw[blue,dash pattern={on 3pt off 1pt},postaction={decoration={markings,mark=at position 0.6 with {\arrow{>>>}}},decorate}]  (#1)-- (#2);
}
\newcommand{\RII}[2]{
\draw[blue,dash pattern={on 3pt off 1pt},postaction={decoration={markings,mark=at position 0.5 with {\arrow{>>}}},decorate}]  (#1)-- (#2);
}
\newcommand{\RI}[2]{
\draw[blue,dash pattern={on 3pt off 1pt},postaction={decoration={markings,mark=at position 0.5 with {\arrow{>}}},decorate}]  (#1)-- (#2);
}

\begin{document}

\title{ Electroweak flavour unification }

\authors{Joe Davighi$^{1,2}$\footnote{{\tt joe.davighi@physik.uzh.ch}} and Joseph Tooby-Smith$^{3,4}$\footnote{{\tt j.tooby-smith@cornell.edu}}}
\institution{UZH}{\centerline{$^1$Physik-Institut, Universit\"at Z\"urich, CH 8057 Z\"urich, Switzerland}}
\institution{DAMTP}{\centerline{$^2$DAMTP, University of Cambridge, Wilberforce Road, Cambridge, UK}}\institution{Cornell}{\centerline{$^3$Department of Physics, LEPP, Cornell University, Ithaca, NY 14853, USA}}\institution{Cav}{\centerline{$^4$Cavendish Laboratory, University of Cambridge, J J Thomson Ave, Cambridge, UK}}

\abstract{ 
We propose that the electroweak and flavour quantum numbers of the Standard Model (SM) could be unified at high energies in an $SU(4)\times Sp(6)_L \times Sp(6)_R$ anomaly-free gauge model. All the SM fermions are packaged into two fundamental fields, $\Psi_L \sim (\mathbf{4}, \mathbf{6}, \mathbf{1})$ and $\Psi_R\sim (\mathbf{4}, \mathbf{1},\mathbf{6})$, thereby explaining the origin of three families of fermions. The SM Higgs, being electroweakly charged, necessarily becomes charged also under flavour when embedded in the UV model. It is therefore natural for its vacuum expectation value to couple only to the third family. The other components of the UV Higgs fields are presumed heavy. Extra scalars are needed to break this symmetry down to the SM, which can proceed via `flavour-deconstructed' gauge groups; for instance, we propose a pattern $Sp(6)_L \to \prod_{i=1}^3 SU(2)_{L,i} \to SU(2)_L$ for the left-handed factor. When the heavy Higgs components are integrated out, realistic quark Yukawa couplings with in-built hierarchies are naturally generated without any further ingredients, if we assume the various symmetry breaking scalars condense at different scales. The CKM matrix that we compute is not a generic unitary matrix, but it can precisely fit the observed values.
} 
\date{}
\maketitle

\setcounter{tocdepth}{3}
\tableofcontents

\hypersetup{linkcolor=blue}

\section{Introduction}\label{sec:intro}

The Standard Model (SM) offers an extremely successful description of particle physics phenomena at energies up to 1 TeV or so. But it is rather complicated. The existence of three generations of matter, and the particular structure observed in their masses and mixings, has no explanation. Even for a single generation the structure is elaborate, with 15 Weyl fermions packaged into five irreducible representations (irreps) of a SM gauge symmetry that is not even semi-simple, and which are cutely arranged so that all gauge anomalies cancel.

Unification is the attempt to explain this observed SM structure as a consequence of something simpler at high energies. For a single generation of fermions, the SM embeds snugly inside $SU(5)$~\cite{Georgi:1974sy}, with the 15 SM fermions spread across the ${\bf 10}$ and $\overline{\bf 5}$ representations. Even more strikingly, if a right-handed neutrino is included, the one-generation SM embeds inside $\mathrm{Spin}(10)$~\cite{Fritzsch:1974nn,georgi1975particles} with all fermions packaged into a ${\bf 16}$-dimensional spinor representation.\footnote{By $\SOTen$ we refer to the double cover of $SO(10)$. Note that, while both the groups $\SOTen$ and $SO(10)$ share the same Lie algebra $\so(10)$, only the former has ${\bf 16}$-dimensional  spinor representations.} 
However, in either scenario, flavour remains as mysterious as before; to fit the three generations that we see, the $SU(5)$ ($\mathrm{Spin}(10)$) GUT now needs six (three) irreps -- no longer so neat. Moreover, the quark and lepton masses and mixings remain arbitrary. Since flavour is such a rich source of unexplained structure in the SM, it is intriguing to ask whether flavour can be brought into a unified gauge model.

One way to do this is simply to promote flavour to a horizontal gauge symmetry, wherein the gauge symmetry can be factored as $G\cong G_{\text{vert}} \times G_{\text{horiz}}$. Here $G_{\text{vert}}$ breaks to the SM gauge symmetry, while $G_{\text{horiz}}$ breaks ultimately to nothing, giving rise to heavy gauge bosons that mediate very weak flavour-changing forces. Curiously, 
if we want to fit all 48 Weyl fermions of the SM+3$\nu_R$ into a {\em single} irrep ${\bf R}$ of a unified gauge group $G$ that acts faithfully on matter, assuming $G$ is connected and semi-simple, then the only options 
 are $G=\SOTen \times \{SU(2) \text{~or~} SO(3)\}$ with ${\bf R}=({\bf 16}, {\bf 3})$~\cite{Allanach:2021bfe}.\footnote{Interestingly, there is no simple gauge group in which one can `faithfully embed' the SM+3$\nu_R$ with all 48 Weyls sitting in a single representation -- Ref.~\cite{Allanach:2021bfe} amounts to a proof-by-exhaustion of this fact. Such is life!}   
Both these are horizontal extensions of the $\SOTen$ GUT. Such horizontal gauge symmetries, and their controlled breaking, have been much-used to explain quark mass and mixing hierarchies (see {\em e.g.}~\cite{Grinstein:2010ve}). Despite these successes, horizontal gauge symmetries are not so compelling from the unification perspective, since flavour is not really unified with the existing SM gauge structure at all. 

Nevertheless, there are interesting options for unifying flavour with the SM gauge symmetries in a way that cannot be factorized as $G_{\text{vert}} \times G_{\text{horiz}}$, which have been little explored in the model-building literature to date. Recently, Ref.~\cite{Allanach:2021bfe} classified all semi-simple Lie algebras $\g$ in which the 
SM+3$\nu_R$ gauge algebra can be embedded without needing extra fermions.
This list of algebras enumerates all possible ways in which flavour could be intertwined with the SM gauge interactions, albeit subject to the (not insignificant) assumption that there are no extra fermions beyond those of the SM+3$\nu_R$. 

One might judge the `most unified' of these gauge algebras to be those in which the SM fermions are packaged into the smallest number $N_{\text{rep}}$ of irreps. If we moreover insist that the flavour symmetry does not just act horizontally, thereby removing the $N_{\text{rep}}=1$  options discussed above, then three gauge algebras from~\cite{Allanach:2021bfe} emerge as especially interesting, each with $N_{\text{rep}}=2$. These are $\su (12) \oplus \su(2)_L \oplus \su(2)_R$, $\su (4) \oplus \sp(6)_L \oplus \sp(6)_R$, and $\su (4) \oplus \sp(6)_L \oplus \so(6)_R$.
In each case, the 48 Weyl fermions transform in a pair of bifundamentals ({\em e.g.}
$\Psi_L \sim ({\bf 12},{\bf 2},{\bf 1} )$ and $\Psi_R \sim ({\bf 12}, {\bf 1}, {\bf 2})$ for the first case), where the former (latter) representation contains all 24 left-handed (right-handed) SM+3$\nu_R$ Weyl fermions. 

For each of these gauge algebras $\g$, one can embed the\footnote{We are being a little cavalier in referring to `the' SM gauge group. Technically, the SM could have one of four possible gauge groups, of the form $\left(SU(3)\times SU(2) \times U(1)\right)/\Gamma$ where the discrete group $\Gamma$ is isomorphic to one of $\{1, \Z_2, \Z_3, \Z_6\}$ (see {\em e.g.}~\cite{Tong:2017oea}). It is actually the version of the SM gauge group with $\Gamma \cong \Z_3$ that embeds inside all three Pati--Salam-based groups written in (\ref{eq:GGroups}).
}
SM gauge group $\GSM$ inside a number of corresponding Lie groups $G$ with $\text{Lie}(G)=\g$. Valid choices of group are
\begin{align} \label{eq:GGroups}
G_{\mathrm{CF}}&= SU (12) \times SU(2)_L \times SU(2)_R \qquad &&\text{(colour flavour unification)}, \\
G_{\mathrm{EWF}}&= SU (4) \times Sp(6)_L \times Sp(6)_R \qquad &&\text{(LR electroweak flavour unification)}, \nonumber \\
G_{\mathrm{EWF}}^\prime&= SU (4) \times Sp(6)_L \times SO(6)_R \qquad &&\text{(\cancel{LR} electroweak flavour unification)}, \nonumber
\end{align} 
for which the corresponding fermion representations are properly anomaly-free (of both local and global\footnote{Freedom from global gauge anomalies is rather subtle to prove, and can be checked by computing an appropriate bordism group. We include the relevant calculation for $G_{\mathrm{EWF}}$ in Appendix~\ref{app:global}. The other cases are somewhat similar.
} gauge anomalies). Each of these gauge groups fully intertwines a 3-family flavour symmetry with the SM gauge symmetries in a way that doesn't factorize,\footnote{Equivalently, the generators of flavour `rotations' do not commute with the generators of the extended SM gauge symmetry.} and so any one of them would explain the origin of three families in terms of an underlying gauge symmetry. 

Let us describe these gauge groups in a little more detail.
The first two are flavour enriched extensions of the familiar Pati--Salam group $SU(4)\times SU(2)_L\times SU(2)_R$~\cite{Pati:1974yy}, and share its left-right symmetry. In $G_{\mathrm{CF}}$, flavour is unified with the `lepton-enlarged' $SU(4)$  colour symmetry of Pati--Salam to an $SU(12)$ factor, while in $G_{\text{EWF}}$ flavour is unified with the `custodially-enlarged' $SU(2)_L \times SU(2)_R$ electroweak symmetry. Note that one cannot choose to extend either electroweak $SU(2)_{L,R}$ factor to an $SU(6)_{L,R}$, because that theory would suffer from gauge anomalies for each $SU(6)$ factor; it is for the $Sp$ series of Lie groups, and not $SU$, that the fundamental representation remains free of perturbative gauge anomalies.\footnote{On the other hand, the fundamental representation of any $Sp$ group, like $Sp(2) \cong SU(2)$, suffers from a $\Z_2$-valued global anomaly~\cite{Witten:1982fp} -- see also Appendix~\ref{app:global}. Ref.~\cite{Lohitsiri:2019wpq} also considers the $Sp$ series of Lie groups to be the natural generalisation of the SM's $SU(2)_L$ symmetry. 
} The fact that the 3-family SM field content can be embedded in the group $G_{\mathrm{EWF}}$ was in fact first noticed by Kuo and Nakagawa in 1985~\cite{Kuo:1984md}, following~\cite{Kuo:1984gz},  although its consequences as a UV model were little explored.
Finally, the group $G^\prime_{\mathrm{EWF}}$ also unifies flavour with electroweak symmetry, but not in a left-right symmetric way. Rather, in this case hypercharge $U(1)_Y\cong SO(2)_R$ is extended via the $SO$ series to an $SO(6)_R$ factor.
Each of the gauge groups in (\ref{eq:GGroups}) has a generalisation to an arbitrary number of SM generations.\footnote{
For $n_f$ generations of SM fermions, the relevant embeddings in each case are: (a) colour and complex $SU(n_f)$ flavour symmetry are unified via the natural embedding of $SU(4)\times SU(n_f) \hookrightarrow SU(4n_f)$; (b) $Sp(2)_{L(R)}$ and real $SO(n_f)$ flavour symmetry are unified via the natural embedding $Sp(2)\times SO(n_f) \hookrightarrow Sp(2n_f)$; (c) $SO(2)_R$ hypercharge and real $SO(n_f)$ flavour symmetry are unified via the natural embedding $SO(2)_R \times SO(n_f)\hookrightarrow SO(2n_f)$. See Refs.~\cite{dyn52:MaximalSubgroups,Lorente:1972xw,Kuo:1984md,Allanach:2021bfe}.
}

We here initiate the study of gauge-flavour unified symmetries such as $G_{\text{CF}}$, $G_{\text{EWF}}$ and $G^\prime_{\text{EWF}}$ as viable theories of flavour. To study the Yukawa sector in any of these gauge models, the first step is to embed the SM Higgs in representations `$H_G$' of $G$. Certainly in the case of electroweak flavour unification, this means that the Higgs, being charged under the SM electroweak symmetry, necessarily acquires flavour quantum numbers; the minimal option is to embed the $({\bf 1}, {\bf 2}, {\bf 2})\oplus ({\bf 15}, {\bf 2}, {\bf 2})$ Higgs fields of the one-family Pati--Salam model inside the representations $H_G = ({\bf 1}, {\bf 6}, {\bf 6})\oplus ({\bf 15}, {\bf 6}, {\bf 6})$ of $G_{\text{EWF}}$. Knowing that the SM flavour symmetries are broken only by the Yukawa couplings of the fermions to the Higgs, where they are broken badly, such a flavour-distinguishing Higgs field seems ideally suited to explaining the mass and mixing hierarchies. Indeed, a generic potential for $H_G$ that breaks electroweak symmetry will simultaneously break the flavour symmetries; one naturally expects a Higgs that couples only to one family, which we should take to be the third. Thus, for $G_{\text{EWF}}$ and $G^\prime_{\text{EWF}}$, one swiftly arrives at a model in which only the third family fermions are massive at the renormalisable level (\S \ref{sec:UV}), which is an agreeable zeroth order postdiction.

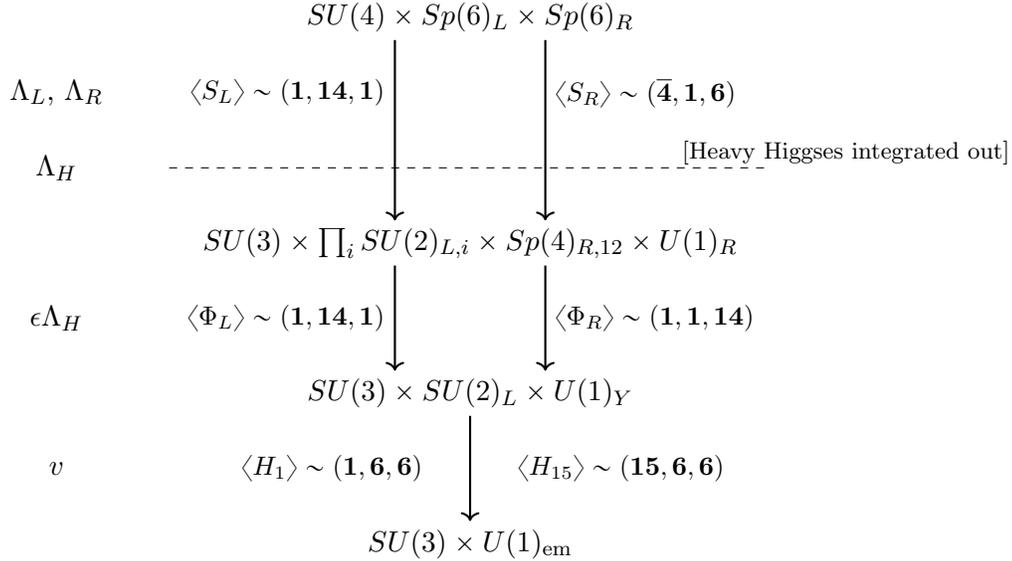
\begin{figure}[t]
\begin{center}
\begin{tikzpicture}
\node at (0,7){$SU(4)\times Sp(6)_L \times Sp(6)_R$};
\node at (0,4){$SU(3)\times \prod_i SU(2)_{L,i} \times Sp(4)_{R,12}\times U(1)_{R}$};
\node at (0,2){$SU(3)\times SU(2)_L \times U(1)_Y$};
\node at (0,0){$SU(3)\times U(1)_{\mathrm{em}}$};
\node at (-5.5,6){$\Lambda_L$, $\Lambda_R$};
\node at (-5.5,5){$\Lambda_H$};
\node at (5,5.2){\footnotesize{[Heavy Higgses integrated out]}};
\node at (-5.5,3){$\epsilon \Lambda_H$};
\node at (-5.5,1){$v$};
\draw[dashed] (-4,5)--(4,5);
\draw[->,thick] (-1,6.7)--(-1,4.3);
\node[anchor=east] at (-1,6){{\small $\langle S_L \rangle \sim ({\bf 1}, {\bf 14}, {\bf 1})$}};
\draw[->,thick] (1,6.7)--(1,4.3);
\node[anchor=west] at (1,6){{\small $\langle S_R \rangle \sim (\overline{\bf 4}, {\bf 1}, {\bf 6})$}};
\draw[->,thick] (-1,3.7)--(-1,2.3);
\node[anchor=east] at (-1,3){{\small $\langle \Phi_L  \rangle \sim ({\bf 1}, {\bf 14}, {\bf 1})$}};
\draw[->,thick] (1,3.7)--(1,2.3);
\node[anchor=west] at (1,3){{\small $\langle \Phi_R \rangle \sim  ({\bf 1}, {\bf 1}, {\bf 14})$}};
\draw[->,thick] (0,1.7)--(0,0.3);
\node[anchor=east] at (-0.5,1){{\small $\langle H_{1} \rangle \sim  ({\bf 1}, {\bf 6}, {\bf 6})$}};
\node[anchor=west] at (0.5,1){{\small $\langle H_{15} \rangle \sim  ({\bf 15}, {\bf 6}, {\bf 6})$}};
\end{tikzpicture}
\end{center}
\caption{The symmetry breaking scheme in our model. At high scales $\Lambda_L$ and $\Lambda_R$ a pair of scalars condenses to break the electroweak-flavour-unified model down to an intermediate gauge theory, which features a deconstructed $SU(2)_L$ symmetry. At a lower scale $\Lambda_H$, the heavy components of the Higgs fields $H_1$ and $H_{15}$ are integrated out. At a slightly lower scale again, indicated by $\epsilon \Lambda_H$, the $\Gint$ theory is broken by the vevs of two more scalars down to the SM. The quantum numbers of all these scalars are recorded in Table~\ref{tab:Scalars}.  } \label{fig:breaking}
\end{figure}

In this paper we focus on the left-right symmetric option $G_{\text{EWF}}$,\footnote{We nonetheless expect that most of our constructions and fermion mass predictions could be adapted to the $G^\prime_{\text{EWF}}$ variant.} and we find that such a gauge theory can provide an elegant explanation of the quark masses and mixing angles observed in Nature. The rough idea is simple, requiring just two symmetry breaking steps and no additional fermions beyond the SM, as follows. 
\begin{itemize}
\item A pair of UV scalar fields $S_L$ and $S_R$ acquire non-zero vacuum expectation values (vevs) at high scales $\Lambda_L$ and $\Lambda_R$ that first break $G_{\mathrm{EWF}}$  down to a family non-universal subgroup $\Gint$, where $\GSM \subset \Gint \subset G_{\text{EWF}}$. (See Fig.~\ref{fig:breaking} for the specific symmetry breaking pattern that we here study.) The vev of the Higgs fields, which start off in representations $H_1\sim({\bf 1}, {\bf 6}, {\bf 6})$ and $H_{15}\sim({\bf 15}, {\bf 6}, {\bf 6})$, emerge in representations of $\Gint$ that couple only to the third family. 
\item The other Higgs components, which couple to lighter families, are presumed heavy and are integrated out at a high scale $\Lambda_H<\Lambda_{L,R}$.
\item Two more scalars $\Phi_L$ and $\Phi_R$ must then acquire vevs at lower scales $\sim \epsilon \Lambda_H$, where $\epsilon<1$ indicates a number of parameters that denote small scale separations needed in the model, which break $\Gint \to \GSM$. Now, by including $G_{\mathrm{EWF}}$-invariant interactions between $H_1$, $H_{15}$, and $\Phi_{L,R}$ in the scalar potential of the UV model, we can compute the tower of higher-dimension operators $\{\mathcal{O}^{d>4}_i\}$ appearing in the $\Gint$-invariant effective field theory (EFT), that are generated by integrating out the heavy Higgs components. Most importantly, there are Yukawa-like operators for the light fermions, involving insertions of components of $\Phi_{L,R}$. 
\end{itemize}
Once $\Phi_{L,R}$ acquire their vevs, the higher dimension operators $\mathcal{O}^{d>4}_i$ match onto Yukawa couplings in the SM, with hierarchies determined by the EFT expansion parameters $\sim \epsilon$. Importantly, we show that there is enough freedom in the model to fit all quark masses and mixing angles to the data (as well as the charged lepton masses -- we postpone a discussion of neutrino mass generation for future work). We emphasize that no additional scalars beyond those strictly necessary to break $\GUV \to \Gint\to \GSM$ are needed, and no extra fermions are needed whatsoever.

The big trade off, of course, like in any GUT (and in many UV models that seek to explain flavour), is that one must swallow a large scalar sector with a tuned potential. Assuming that the extra components of the Higgs fields are heavy is reminiscent of the well-known `doublet-triplet problem' that afflicts GUTs, although the problem is less severe here because the extra states do not lead to proton decay -- this also means the high scale $\Lambda_H$ can be much lower than the traditional GUT scale. Moreover, to generate the mass hierarchies requires a small separation of the two symmetry breaking scales by an order of magnitude or two. Recent work~\cite{Allwicher:2020esa} at least suggests that maintaining such a separation of scales, in the context of flavour model building, can be radiatively stable. And, of course, the presence of many heavy scalars that couple to the Higgs would exacerbate the electroweak hierarchy problem -- which is essentially an unavoidable problem in models with enlarged symmetries in the UV.
All of these requirements place constraints on the coefficients appearing in the scalar potential of the UV model, a detailed study of which is beyond the scope of this paper. 

Finally, we remark that a further hint for high scale gauge-flavour unification comes from an idea in quantum gravity, which suggests that there are no global symmetries in the UV; either a symmetry is gauged, as would be the fate of the SM's flavour symmetries if there were high scale electroweak flavour unification, or it is explicitly broken. Well known arguments from black hole heuristics gave rise to this idea (going back to~\cite{Hawking:1975vcx}), which has subsequently been proven in the context of holography~\cite{Harlow:2018tng,Harlow:2018jwu}, and in perturbative string theory~\cite{Banks:2010zn}. 

The structure of this paper is as follows. In \S \ref{sec:UV} we review some basic facts about the $Sp(6)$ Lie group and introduce some helpful notation, before setting out the symmetries and couplings in the UV model. In \S \ref{sec:breaking} we explain the symmetry breaking pattern in detail. We show in detail how the SM quark masses and mixings are generated in \S \ref{sec:yukawa}. Finally, we conclude and discuss some interesting future directions in \S\ref{sec:Conclusion}.

\section{Electroweak flavour unification at high energies} \label{sec:UV}

In this paper we propose a gauge-flavour unified model based on the UV gauge group
\be \label{eq:UVgaugegroup}
\GUV := SU(4) \times Sp(6)_L \times Sp(6)_R\, .
\ee
Since the Lie group $Sp(6)$ may be unfamiliar to some model-builders, we begin by recalling its definition\footnote{Since there are a number of conventions in use for the symplectic groups, we emphasize that in our convention $Sp(2)\cong SU(2)$, with $Sp(2N)$ having a fundamental representation of complex dimension $2N$.
} and setting out some useful conventions.

\subsection{Notation and conventions} \label{sec:notation}

The matrix group $Sp(6)$ consists of $6\times 6$ special unitary matrices $U$ such that 
$U^T\Omega U=\Omega$, where $\Omega=\begin{psmallmatrix} 0 & I_3 \\ -I_3 & 0 \end{psmallmatrix}$.
The Lie algebra of $Sp(6)$, denoted $\mathfrak{sp}(6)$, is
\begin{align}
\mathfrak{sp}(6):=\{X\in M_{6\times 6}(\C)\mid \Omega X=-X^T \Omega, X=X^\dagger, \mathrm{Tr}(X)=0\}\, .
\end{align}
The dimension of $\mathfrak{sp}(6)$ is 21. 

We denote by $\{a_1,a_2,a_3,a_4\}$ the standard basis for the vector space $V_4 \cong \C^4$ acted on by the fundamental representation ${\bf 4}$ of $SU(4)$, and the basis for the conjugate representation ${\bf \bar{4}}$ by $\{a_1^\ast, a_2^\ast, a_3^\ast,a_4^\ast\}$. The basis for the vector space $V_L \cong \C^6$ acted on by the fundamental representation ${\bf 6}$ of $Sp(6)_L$ is denoted $\{b_1,b_2,b_3,b_4,b_5,b_6\}$, and the basis for $V_R \cong \C^6$ acted on by the fundamental of $Sp(6)_R$ is denoted $\{c_1,c_2,c_3,c_4,c_5,c_6\}$. For example, in this notation the matrix $\Omega$ above, for $Sp(6)_L$, takes the form
\begin{align}
\Omega=b_1^\dagger \wedge b_4^\dagger+b_2^\dagger \wedge b_5^\dagger+b_3^\dagger \wedge b_6^\dagger,
\end{align}
where $\{b_i^\dagger\}$ is the dual basis to $\{b_i\}$.
Complex conjugation $(\ast)$ is defined throughout to be the complex-linear map $a_i\mapsto a_i^\ast$, $b_i\mapsto \Omega_{ji}b_{j}$ and $c_i\mapsto \Omega_{ji}c_{j}$. A real field is one for which $\phi=\phi^\ast$. With the exception of Weyl fermions, we use a `bar' (\emph{e.g.} $\bar\phi$) to indicate a distinct symbol, not complex conjugation. We will use it, nevertheless, such that if a reality 
condition is imposed on the relevant object then $\bar\phi=\phi^\ast$.

\paragraph{Inner products:}
In what follows, we make use of two inner products. The first, which we denote $\langle\cdot,\cdot\rangle_1$, is an inner product on the complex vector space $V_L\otimes V_R$, defined as 
\begin{align} \label{eq:innerP_1}
\langle A,B\rangle_1=\Tr(\Omega^T A^\dagger \Omega B)\, .
\end{align}
The second, which we denote $\langle\cdot,\cdot\rangle_{15}$, is an inner product on $V_{4}\otimes V_{4}^\ast \otimes V_L\otimes V_R$, defined as 
\begin{align} \label{eq:innerP_15}
\langle A,B\rangle_{15}=\sum_{ij} \Tr(\Omega^T A_{ij}^\dagger\Omega B_{ji})\, ,
\end{align}
where the sum is over $SU(4)$ indices, and the $\Tr$is over the $Sp(6)$ indices.  The inner products satisfy the additional relations 
\begin{align}
\langle A,B\rangle_a=\langle B^\ast,A^\ast\rangle_a\, , \qquad a \in \{1,15\}.
\end{align}

\paragraph{Diagrams:}
It will be convenient when we come to draw Feynman diagrams to introduce a pictorial notation for the `flow' of $Sp(6)$ indices.
We thus introduce solid red lines marked with 1, 2, or 3 arrowheads to denote specific contractions of $Sp(6)_L$ fundamental representations $x$ and $y$, as follows:
\begin{align}
\begin{tikzpicture}
\begin{feynman}
\vertex (a);
\vertex [right=1in of a] (b);
\node at (a) [label=left:{ $\big[x$}]{};
\node at (b) [label=right:{$y\big]$}]{};
\begin{scope}[red,decoration={
    markings,
    mark=at position 0.6 with {\arrow{>}}}
    ] 
	\draw[postaction={decorate}] plot [smooth] coordinates {(a) (b)};
	\end{scope}
\end{feynman}
\node [below right=-0.07in and 0.2in of b] (c) {$\sim x_1  y_4,\quad$};
\end{tikzpicture}
\begin{tikzpicture}
\begin{feynman}
\vertex (a);
\vertex [right=1in of a] (b);
\node at (a) [label=left:{ $\big[x$}]{};
\node at (b) [label=right:{$y\big]$}]{};
\begin{scope}[red,decoration={
    markings,
    mark=at position 0.6 with {\arrow{<}}}
    ] 
	\draw[postaction={decorate}] plot [smooth] coordinates {(a) (b)};
	\end{scope}
	\node [below right=-0.07in and 0.2in of b] (c) {$\sim x_4  y_1,$};
\end{feynman}
\end{tikzpicture}
\nonumber\\
\begin{tikzpicture}
\begin{feynman}
\vertex (a);
\vertex [right=1in of a] (b);
\node at (a) [label=left:{ $\big[x$}]{};
\node at (b) [label=right:{$y\big]$}]{};
\begin{scope}[red,decoration={
    markings,
    mark=at position 0.6 with {\arrow{>>}}}
    ] 
	\draw[postaction={decorate}] plot [smooth] coordinates {(a) (b)};
	\end{scope}
	\node [below right=-0.07in and 0.2in of b] (c) {$\sim x_2  y_5,\quad$};
\end{feynman}
\end{tikzpicture}
\begin{tikzpicture}
\begin{feynman}
\vertex (a);
\vertex [right=1in of a] (b);
\node at (a) [label=left:{ $\big[x$}]{};
\node at (b) [label=right:{$y\big]$}]{};
\begin{scope}[red,decoration={
    markings,
    mark=at position 0.6 with {\arrow{<<}}}
    ] 
	\draw[postaction={decorate}] plot [smooth] coordinates {(a) (b)};
	\end{scope}
	\node [below right=-0.07in and 0.2in of b] (c) {$\sim x_5  y_2,$};
\end{feynman}
\end{tikzpicture}
\nonumber \\
\begin{tikzpicture}
\begin{feynman}
\vertex (a);
\vertex [right=1in of a] (b);
\node at (a) [label=left:{ $\big[x$}]{};
\node at (b) [label=right:{$y\big]$}]{};
\begin{scope}[red,decoration={
    markings,
    mark=at position 0.6 with {\arrow{>>>}}}
    ] 
	\draw[postaction={decorate}] plot [smooth] coordinates {(a) (b)};
	\end{scope}
	\node [below right=-0.07in and 0.2in of b] (c) {$\sim x_3  y_6,\quad$};
\end{feynman}
\end{tikzpicture}
\begin{tikzpicture}
\begin{feynman}
\vertex (a);
\vertex [right=1in of a] (b);
\node at (a) [label=left:{ $\big[x$}]{};
\node at (b) [label=right:{$y\big]$}]{};
\begin{scope}[red,decoration={
    markings,
    mark=at position 0.6 with {\arrow{<<<}}}
    ] 
	\draw[postaction={decorate}] plot [smooth] coordinates {(a) (b)};
	\end{scope}
	\node [below right=-0.07in and 0.2in of b] (c) {$\sim x_6  y_4.$};
\end{feynman}
\end{tikzpicture}
\end{align} 
The significance of the number of arrowheads is that it will match the family index of the SM fermion. We introduce analogous lines for $Sp(6)_R$, with the exception that solid red lines $\left(\begin{tikzpicture}
\begin{feynman}
\vertex (a);
\vertex [right=0.3in of a] (b);
\node at (a) []{};
\node at (b) []{};
\begin{scope}[red,decoration={
    markings,
    mark=at position 0.6 with {\arrow{}}}
    ] 
	\draw[postaction={decorate}] plot [smooth] coordinates {(a) (b)};
	\end{scope}
\end{feynman}
\end{tikzpicture}\right)$
are replaced by dashed blue lines
$\left(\begin{tikzpicture}
\begin{feynman}
\vertex (a);
\vertex [right=0.3in of a] (b);
\node at (a) []{};
\node at (b) []{};
\begin{scope}[blue,dash pattern={on 3pt off 1pt},decoration={
    markings,
    mark=at position 0.6 with {\arrow{}}}
    ] 
	\draw[postaction={decorate}] plot [smooth] coordinates {(a) (b)};
	\end{scope}
\end{feynman}
\end{tikzpicture}\right)$.

\subsection{Embedding the SM particles}

We now describe the basic elements of the model.
Firstly, we take the SM gauge group to be
\be \label{eq:SMZ3}
\GSM:=\frac{SU(3)\times SU(2)_L \times U(1)_Y}{\Z_3} \, .
\ee
Here, the $\Z_3$ quotient is generated by the element $(\omega, {\bf 1}, e^{2\pi i /3})\in SU(3)\times SU(2)_L \times U(1)_Y$, where $\omega$ is the generator of the $\Z_3$ centre of $SU(3)$ such that $\omega^3 = {\bf 1} \in SU(3)$.
This version of the SM gauge group embeds inside the UV gauge group
(\ref{eq:UVgaugegroup}).\footnote{
To see that it is the particular group (\ref{eq:SMZ3}) that embeds inside $\GUV$, first consider the map $\beta:SU(3)\times SU(2)_L \times U(1)_Y \to SU(4)\times SU(2)_L\times SU(2)_R:(h,g_L,\alpha) \mapsto \left( \begin{psmallmatrix} \alpha h & 0\\ 0 & \alpha^{-3}\end{psmallmatrix}, g_L,  \begin{psmallmatrix} \alpha^3 & 0\\ 0 & \alpha^{-3}\end{psmallmatrix}
  \right)$. This map is not injective and so not an embedding~\cite{Baez:2009dj}. Rather, $\ker \beta \cong \Z_3$, generated by the element $\left( e^{-2\pi i/3} \mathbf{1}_3, 1,  e^{2\pi i/3}\right)$. Quotienting by $\ker \beta$, we arrive at an injective map $\GSM \hookrightarrow SU(4)\times SU(2)_L \times SU(2)_R$, which can be composed with the injection $SU(4)\times SU(2)_L \times SU(2)_R \hookrightarrow \FSS$ to embed $\GSM$ in $\GUV$.
}
As described in the Introduction, the $\GUV$ symmetry unifies the SM electroweak gauge symmetries with the SM flavour symmetries that act on the 3 generations of matter, while simultaneously unifying quarks and leptons via the enlarged colour group $SU(4)$.

All the eighteen fermion multiplets of the SM, including three right-handed neutrinos $\nu_{R,f}$, are packaged together into just two Weyl fermion fields, 
\begin{nalign}
\Psi_L \sim ({\bf 4}, {\bf 6}, {\bf 1}) 
&= Q^{i,k}_f a_i\otimes b_{3k+f}+L^{k}_f a_4 \otimes b_{3k+f} \\
&=D^{i}_{L,f} a_i \otimes b_{f}+U^i_{L,f} a_i\otimes b_{3+f}+E_{L,f} a_4\otimes b_f+\nu_{L,f} a_4\otimes b_{3+f}\, , \\
\Psi_R \sim ({\bf 4}, {\bf 1}, {\bf 6}) 
&=D^{i}_{R,f} a_i \otimes c_{f}+U^i_{R,f} a_i\otimes c_{3+f}+E_{R,f} a_4\otimes c_f+\nu_{R,f} a_4\otimes c_{3+f}\, ,
\end{nalign}
each of which has 24 components. The index $i\in\{1,2,3\}$ labels colour, with the fourth component of $a_i$ being reserved for leptons, $k\in\{0,1\}$ labels $SU(2)_L$ isospin, and $f\in\{1,2,3\}$ labels the family.

The SM Higgs field is embedded in a pair of complex UV scalar fields,
\be
H_{1} \sim ({\bf 1}, {\bf 6}, {\bf 6}), \qquad H_{15} \sim ({\bf 15}, {\bf 6}, {\bf 6}), 
\ee
with which we write down Yukawa couplings in the renormalisable UV theory, schematically
\be 
\mathcal{L}_{\mathrm{yuk}} = y_{1} \overline{\Psi}_L H_1 \Psi_R +  y_{15} \overline{\Psi}_L H_{15} \Psi_R +\overline{y}_1 \overline{\Psi}_L H_1^\ast \Psi_R+  \overline{y}_{15} \overline{\Psi}_L H_{15}^\ast \Psi_R + \mathrm{h.c.}
\ee
Precisely, we can use the inner products (\ref{eq:innerP_1}) and (\ref{eq:innerP_15}) to indicate how the group indices are contracted:
\begin{align} \label{eq:ren_yuk}
\mathcal{L}_{\mathrm{yuk}} = \sum_{a\in\{1,15\}} \bigg\{ y_a \langle (\Gamma_a\Psi_{R}\overline{\Psi}_{L} )^\dagger,H_a\rangle_{a}+\overline{y}_a\langle H_a, (\Gamma_a\Psi_{R}\overline{\Psi}_{L} )^T\rangle_{a}\\
\nonumber
+y_a^\ast \langle H_a, (\Gamma_a\Psi_{R}\overline{\Psi}_{L} )^\dagger\rangle_{a}+\overline{y}_a^\ast\langle(\Gamma_a\Psi_{R}\overline{\Psi}_{L} )^T,H_a\rangle_{a} \bigg\}\, ,
\end{align}
where the $\Gamma_1=\mathrm{Tr}_4$ indicates a trace of $SU(4)$ indices and $\Gamma_{15}$ is the identity. 

Since the Higgs fields transform in the bifundamental representation of the flavour-enriched electroweak symmetry $Sp(6)_L \times Sp(6)_R$, a generic electroweak symmetry breaking vev will also pick up a direction in flavour space. 
The vev directions can be such that the renormalisable Yukawa couplings (\ref{eq:ren_yuk}) only give masses to one family, which we are free to identify with the third family. Specifically, the Higgs vevs are
\begin{align}
\langle H_1\rangle &= v_1 b_3\otimes c_6-\overline{v}_1 b_6\otimes c_3 \, , \label{eq:H1vev} \\
\langle H_{15}\rangle &= \left(
a_1 \otimes a_1^\ast+
a_2 \otimes a_2^\ast+
a_3 \otimes a_3^\ast -3
a_4 \otimes a_4^\ast \right) \otimes \left( v_{15} b_3\otimes c_6- \overline{v}_{15}b_6\otimes c_3\right) \, \label{eq:H15vev} ,
\end{align}
where $v_1$, $\overline{v}_1$, $v_{15}$, and $\overline{v}_{15}$ are four  independent scales.
(If one wants to force $H_a$ to be real, then take $\overline{v}_a=v^\ast_a$.)

Thus, in the renormalisable UV theory, only the third family fermions are massive. To leading order, their masses are 
\begin{nalign} \label{eq:3rd-fam-masses}
m_t &\approx (y_{1} \overline{v}_1 +\overline{y}_1 v_1^\ast)+ (y_{15} \overline{v}_{15} +\overline{y}_{15} v_{15}^\ast), \\
m_b &\approx (y_{1} {v}_1 +\overline{y}_1 \overline{v}_1^\ast)+ (y_{15} {v}_{15} +\overline{y}_{15} \overline{v}_{15}^\ast), \\
m_\tau &\approx(y_{1} {v}_1 +\overline{y}_1 \overline{v}_1^\ast)-3 (y_{15} {v}_{15} +\overline{y}_{15} \overline{v}_{15}^\ast), \\ 
\end{nalign}
mimicking the mass formulae of the one-family Pati--Salam model. 
In this model, unlike a flavour-blind Pati--Salam model, the light families are strictly massless at the renormalisable level. In \S \ref{sec:yukawa} we will see how the light masses can be generated at a lower scale by higher-order operators in the effective theory.

\begin{figure}
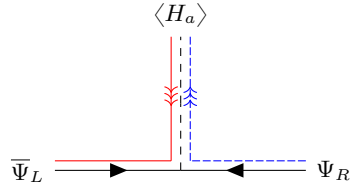

\begin{center}
\colordiagram{
    \fermions
    \HiggsII{L1}{R1}
    \LIII{L0}{L1}
    \RIII{R0}{R1}
}
\end{center}
\caption{The vevs (\ref{eq:H1vev}--\ref{eq:H15vev}) of the Higgs fields $H_1$ and $H_{15}$ couple only to the third family fermions, because of their direction in $Sp(6)_L \times Sp(6)_R$ space. \label{fig:Y3} }
\end{figure}

In principle, one neutrino also acquires a renormalisable Dirac mass, but we assume that a form of seesaw mechanism sends the physical neutrino masses down to the eV scale. We postpone a discussion of neutrino masses for future work.

\section{Symmetry breaking pattern} \label{sec:breaking}

We next discuss the breaking of the UV gauge symmetry $G_{\mathrm{EWF}}$ down to $\GSM$. This requires many scalar fields, and we choose what we believe to be an almost-minimal set of scalars that will do the job -- see Table~\ref{tab:Scalars}. We suppose that these scalars acquire their vevs at different energy scales, resulting in a sequential breaking of $\GUV$ down to $\GSM$ via an intermediate effective field theory (EFT) that we describe in this Section.

\subsection{Flavour deconstruction in the intermediate EFT } \label{sec:breaking1}

The group $\FSS$ has many subgroups that contain $\GSM$, and so there are many possible paths by which one can break $\GUV \to \GSM$. That said, the breaking of the $SU(4)$ colour factor is essentially constrained to be as it is in the one-family Pati--Salam case, {\em i.e.} via $SU(3) \times U(1)_{B-L}$ where $U(1)_{B-L}$ must ultimately combine with a $U(1)$ subgroup of $Sp(6)_R$ to give hypercharge. Thus, the freedom we have in breaking $\GUV \to \GSM$ comes from the plethora of subgroups of the family-enriched $Sp(6)_L \times Sp(6)_R$ symmetry.

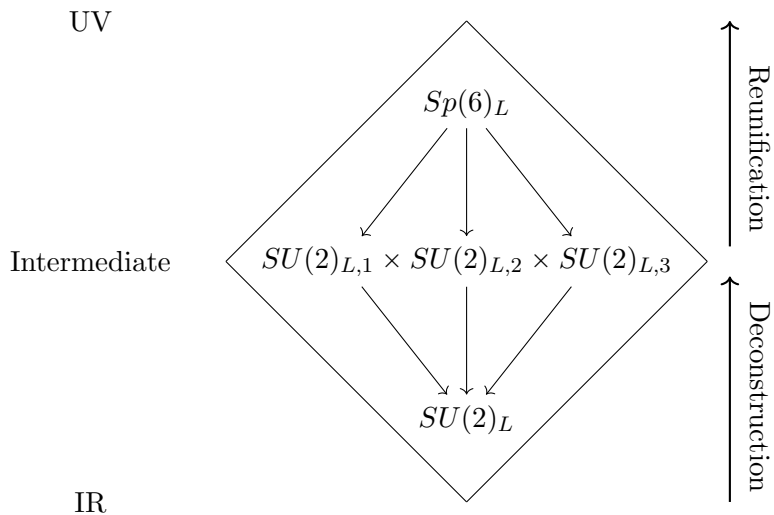
\begin{figure}[h]
\begin{center}
\begin{tikzpicture}
\coordinate (b) at (0,0);
\coordinate (t) at (0,6.4);
\coordinate (l) at (-3.2,3.2);
\coordinate (r) at (3.2, 3.2); 
\draw (b)--(l);
\draw (b)--(r);
\draw (t)--(l);
\draw (t)--(r);
\node (int) at (0,3.2){$SU(2)_{L,1}\times SU(2)_{L,2}\times SU(2)_{L,3}$};
\node (int2) at (1.5,3.4){};
\node (int3) at (-1.5,3.4){};
\node (int2b) at (1.5,3.0){};
\node (int3b) at (-1.5,3.0){};
\node (su2)  at (0,1.1){$SU(2)_L$};
\node (sp6) at (0,5.3){$Sp(6)_L$};
\draw[->] (sp6)--(int);
\draw[->] (sp6) --  (int2);
\draw[->] (sp6) --  (int3);
\draw[->] (int)--(su2);
\draw[->] (int2b)--(su2);
\draw[->] (int3b)--(su2);
\coordinate (decb) at (3.5,0);
\coordinate (dect) at (3.5,3.0);
\node[rotate=270] at (3.9,1.4){Deconstruction};
\draw[->,thick] (decb)--(dect);

\coordinate (ret) at (3.5,6.4);
\coordinate (reb) at (3.5,3.4);
\node[rotate=270] at (3.9,4.7){Reunification};
\draw[->,thick] (reb)--(ret);
\node at (-5.,0){IR};
\node at (-5.,3.2){Intermediate};
\node at (-5.,6.4){UV};
\end{tikzpicture}
\end{center}
\caption{The universal $SU(2)_L$ weak interactions that we see at low energies could be deconstructed according to flavour at higher energies, before being ultimately re-unified with flavour in the UV, via an $Sp(6)_L$ symmetry. \label{fig:deconstruction} }
\end{figure}

One intriguing possibility, which is the route we explore in this paper, exploits the fact that $Sp(6)_L$ contains a `flavour-deconstructed' electroweak symmetry group,
$SU(2)_{L,1} \times SU(2)_{L,2} \times SU(2)_{L,3} \subset Sp(6)_L $,\footnote{The same breaking pattern, $Sp(2n_f) \to \prod_{i=1}^{n_f} SU(2)_{L,i} \to SU(2)_L$, can be realised for any number $n_f$ of SM generations.} 
where $SU(2)_{L,i}$ denotes the usual $SU(2)_L$ factor of the SM but acting only on the $i^\mathrm{th}$ family. Breaking $Sp(6)_L$ down to this subgroup can be achieved using a scalar field in the antisymmetric 2-index ${\bf 14}$, the joint-smallest dimension irrep of $Sp(6)_L$ after the fundamental. We remark that Ref.~\cite{Kuo:1984md} made use of a similar symmetry breaking pattern.

We thus suppose that at a high scale $\Lambda_L$ a real scalar field $S_{L}\sim ({\bf 1},{\bf 14},{\bf 1})$ acquires the vev
\be \label{eq:SLvev}
\langle S_L \rangle = \Lambda_L \left(b_1\wedge b_4 - b_3\wedge b_6 \right)\, ,
\ee
where we use `$\wedge$' to denote the antisymmetrization over $Sp(6)_L$ fundamental indices.\footnote{In fact, one achieves the same symmetry breaking pattern for any non-zero vev in the vector space spanned by (\ref{eq:SLvev}) and $b_3\wedge b_6-b_2\wedge b_5$, so this is a rather generic (as well as minimal) breaking pattern.}
This induces the symmetry breaking 
\be \label{eq:Lbreaking}
Sp(6)_L \to SU(2)_{L,1} \times SU(2)_{L,2} \times SU(2)_{L,3} \, .
\ee 
Thus, at intermediate energies, each SM family interacts with its own set of  $SU(2)_{L,i}$ weak gauge bosons. 
There are 12 broken generators, and the corresponding heavy gauge bosons consist of
three sets of flavour-changing $SU(2)_L$ triplets $W^{\prime a}$, plus
three additional SM singlet $\ZP$ bosons.


As an aside, we remark that similar `flavour-deconstructed' SM gauge symmetries have recently been used as part of a larger `Pati--Salam cubed' symmetry $\prod_{i=1}^3 SU(4)_i \times SU(2)_{L,i} \times SU(2)_{R,i}$~\cite{Bordone:2017bld,Bordone:2018nbg,Fuentes-Martin:2020pww}, to explain both fermion masses and the $B$ physics anomalies recently measured by LHCb. In these works, a fifth dimension punctuated by 4d matter branes is a possible origin of such flavour deconstruction. Here we see that deconstruction of $SU(2)_L$ could alternatively emerge from a 4d gauge model in which flavour and $SU(2)$ symmetry are eventually re-unified deeper in the UV.\footnote{We emphasize that a flavour-deconstruction of the $SU(4)$ factor is needed if we want to explain the $B$ physics anomalies in $\mathrm{PS}^3$ models, since one needs a $U_1$ leptoquark coupled predominantly to the third family. }
We summarize this qualitative picture in the cartoon in Fig.~\ref{fig:deconstruction}.

\begin{table}
\begin{center}
\begin{tabular}{|c|c||c|c|}
\hline
UV fermion & Rep & Intermediate fermion & Rep\\
\hline
$\Psi_L$ & $(\mathbf{4,6,1})$  & $Q_1$ &
$[\mathbf{3,(2,1,1),1}]_{1}$ \\
& &$Q_2$ &
$[\mathbf{3,(1,2,1),1}]_{1}$ \\
& & $Q_3$ &
$[\mathbf{3,(1,1,2),1}]_{1}$ \\
& & $L_1$ &
$[\mathbf{1,(2,1,1),1}]_{-3}$ \\
& & $L_2$ &
$[\mathbf{1,(1,2,1),1}]_{-3}$ \\
& & $L_3$ & $[\mathbf{1,(1,1,2),1}]_{-3}$ \\
\hline
$\Psi_R$ &
$(\mathbf{{4},1,6})$ & $\mathcal{Q}_{R,12}$ & $[\mathbf{{3},(1,1,1),4}]_{1}$\\
& & $\mathcal{L}_{R,12}$ & $[\mathbf{1,(1,1,1),4}]_{-3}$\\
& & $U_3$ & $[\mathbf{{3},(1,1,1),1}]_{4}$\\
& & $D_3$ & $[\mathbf{{3},(1,1,1),1}]_{-2}$\\
& & $E_3$ & $[\mathbf{1,(1,1,1),1}]_{-6}$\\
& & $\nu_{R,3}$ & $[\mathbf{1,(1,1,1),1}]_{0}$\\
\hline
\end{tabular}
\end{center}
\caption{Representations of the UV fermions under $\GUV=\FSS$ (left), and how these decompose under the symmetry breaking $\GUV \to \GintE$ that occurs at the high scale $\Lambda_H$.  } \label{tab:Matter_Content}
\end{table}

We must also decide how to break the $Sp(6)_R$ symmetry. 
One well-motivated option is to break $Sp(6)_R$ alongside $SU(4)$ in such a way that the first two families of right-handed fields remain unified. (The breaking of degeneracy in the 1-2 Yukawa sector will then arise in the next symmetry breaking step {\em i.e.} the breaking down to the SM.)
To achieve this, the minimal choice is to take a complex scalar field $S_R \sim (\overline{\bf 4}, {\bf 1}, {\bf 6})$ which also gets a vev at a high scale $\Lambda_R$, which need not coincide with $\Lambda_L$: 
\be \label{eq:SRvev}
\langle S_R \rangle = \Lambda_R\,  a_4^\ast \otimes c_3\, .
\ee
This vev triggers the high-scale symmetry breaking
\be \label{eq:Rbreaking}
SU(4)\times Sp(6)_R\to SU(3)\times Sp(4)_{R,12}\times U(1)_R\, ,
\ee
where the right-handed fermions in the first and second family remain packaged into fundamental reps of $Sp(4)_{R,12}$.
Here $U(1)_R$ acts as hypercharge on the third family and on the left-handed fermions; for the light right-handed fermions, SM hypercharge will emerge as a linear combination of $U(1)_R$ and a $U(1)$ subgroup of $Sp(4)_{R,12}$. 
A total of 17 heavy gauge bosons result from this breaking: one $U_1$ vector leptoquark, in the representation $(\bar{\bf 3}, {\bf 1})_{-4}$, three charged $Z^\pm$ bosons, that is, complex vector bosons in the representation $({\bf 1}, {\bf 1})_{6}$, and five (real) $\ZP$ bosons.

Putting (\ref{eq:Lbreaking}) and (\ref{eq:Rbreaking}) together, the UV symmetry $\GUV$ is broken at high scales $\Lambda_{L,R}$ by $\langle S_L \rangle$ and $\langle S_R \rangle$ down to an intermediate gauge symmetry,
\be
\GUV \xrightarrow{\Lambda_L,\, \Lambda_R} \Gint := SU(3)\times \prod_i SU(2)_{L,i} \times Sp(4)_{R,12}\times U(1)_R\, .
\ee
The decomposition of the SM fermion fields under $\Gint$ is recorded in Table~\ref{tab:Matter_Content}.
The Higgs fields $H_1$ and $H_{15}$ decompose into a multitude of scalars under $\Gint$, {\em viz.}
\begin{align} \label{eq:Hdecomp}
H_{1,15}\mapsto 
&[{\bf 1},({\bf 2},{\bf 1},{\bf 1}),{\bf 1}]_{-3}\oplus [{\bf 1},({\bf 1},{\bf 2},{\bf 1}),{\bf 1}]_{-3}\oplus \underline{[{\bf 1},({\bf 1},{\bf 1},{\bf 2}),{\bf 1}]_{-3}} \nonumber\\
&\oplus [{\bf 1},({\bf 2},{\bf 1},{\bf 1}),{\bf 1}]_{3}\oplus [{\bf 1},({\bf 1},{\bf 2},{\bf 1}),{\bf 1}]_{3}\oplus \underline{[{\bf 1},({\bf 1},{\bf 1},{\bf 2}),{\bf 1}]_{3}}  \nonumber\\
&\oplus [{\bf 1},({\bf 2},{\bf 1},{\bf 1}),{\bf 4}]_{0}\oplus [{\bf 1},({\bf 1},{\bf 2},{\bf 1}),{\bf 4}]_{0} \oplus[{\bf 1},({\bf 1},{\bf 1},{\bf 2}),{\bf 4}]_{0}\, \nonumber \\
&\oplus \, \{\text{$SU(3)$ triplets and octets for $H_{15}$}\}\, .
\end{align}
The underlined components, which we name
\begin{align}
\overline{\mathcal{H}}_1 &\sim [{\bf 1},({\bf 1},{\bf 1},{\bf 2}),{\bf 1}]_{-3}, \quad &&\mathcal{H}_1 \sim [{\bf 1},({\bf 1},{\bf 1},{\bf 2}),{\bf 1}]_{+3} \, , \\
\overline{\mathcal{H}}_{15} &\sim [{\bf 1},({\bf 1},{\bf 1},{\bf 2}),{\bf 1}]_{-3}, \quad &&\mathcal{H}_{15} \sim [{\bf 1},({\bf 1},{\bf 1},{\bf 2}),{\bf 1}]_{+3} \, ,
\end{align}
will contain the physical SM Higgs doublet; these components remain light and acquire the EWSB vev (\ref{eq:H1vev}--\ref{eq:H15vev}).
By construction, they are charged only under the third family symmetry factors $SU(2)_{L,3}$ and $U(1)_R$.

The other Higgs components written in (\ref{eq:Hdecomp}), {\em i.e.} those not underlined, do not acquire vevs, and are assumed to be heavy with masses \be
M_{H} \approx \Lambda_H<\Lambda_{L,R},
\ee
where $\Lambda_H$ defines an EFT matching scale at which the heavy Higgses are integrated out. From their representations under the flavour-deconstructed $\Gint$ symmetry, we see that there are Higgs components that couple to each pair of SM families (one left-handed, one right-handed); this will be important in \S \ref{sec:yukawa}.

\begin{table}
\begin{widepageIII}
\begin{center}
\begin{tabular}{|c|c|c|c|}
\hline
 & $\GUV$ irrep & Vev direction & $\Gint$ irrep(s) of vev \\
\hline
$S_L$ & $(\mathbf{1,14,1})$& $b_1\wedge b_4 -b_3\wedge b_6$ & NA\\
$S_R$ & $(\mathbf{\overline{4},1,6})$& $a_4^\ast \otimes c_3$ & NA\\
\hline\hline
$ \Phi_L$ & $(\mathbf{1,14,1})$& $\epsilon_L^{12} \left(b_1\wedge b_5 +b_2\wedge b_4\right)+\epsilon_L^{23} \left(b_2\wedge b_6 +b_3\wedge b_5\right)$ & $\phi_L^{12}\sim[\mathbf{1,(2,2,1),1}]_{0}$,\\
 & & & $\phi_L^{23}\sim[\mathbf{1,(1,2,2),1}]_{0}$\\
 \hline
$\Phi_{R}$ & $(\mathbf{1,1,14})$& $\epsilon_R^{23}(w_{23}c_2\wedge c_6+\overline{w}_{23}c_3\wedge c_5)$ & $\phi_{R}^{23}\sim[\mathbf{1,(1,1,1),4}]_{+3}$,\\
& & $+\epsilon_{R}^{12}(w_{12} c_1\wedge c_5+\overline{w}_{12} c_2 \wedge c_4)$ & $\overline{\phi}_R^{23}\sim[\mathbf{1,(1,1,1),4}]_{-3}$,\\
& & & $\phi_R^{12}\sim[\mathbf{1,(1,1,1),5}]_{0}$\\
\hline \hline
$H_1$ & $(\mathbf{1,6,6})$& $v_1 b_3\otimes c_6-\overline{v}_1 b_6\otimes c_3$ & $\mathcal{H}_{a}\sim[\mathbf{1,(1,1,2),1}]_{+3}$,\\
$H_{15}$ & $(\mathbf{15,6,6})$& $(a_i\otimes a_i^\ast-3 a_4\otimes a_4^\ast)\otimes(v_{15} v_3\otimes c_6-\overline{v}_{15} b_6\otimes c_3)$ &
 $\overline{\mathcal{H}}_{a}\sim[\mathbf{1,(1,1,2),1}]_{-3}$\\
\hline
\end{tabular}
\end{center}
\end{widepageIII}
\caption{The set of scalar fields needed to break the UV symmetry $\FSS$ eventually down to $SU(3)_C \times U(1)_\text{em}$. We record the directions of the corresponding vevs in $\FSS$ space, where $a_i$, $b_i$, and $c_i$ index the fundamental representations of $SU(4)$, $Sp(6)_L$, and $Sp(6)_R$ respectively, as well as the representations of the intermediate gauge symmetry $\GintE$ in which the vevs sit. 
} \label{tab:Scalars}
\end{table}

\subsection{Breaking to the SM} \label{sec:breaking2}

The intermediate gauge symmetry $\Gint$ must be broken down to $\GSM$, and this occurs at energy scales below $\Lambda_H$. 
We refer the reader back to Fig.~\ref{fig:breaking}, which summarizes the overall symmetry breaking scheme for our model.

A strikingly minimal sector consisting of two scalars $\Phi_L \sim (\mathbf{1},\mathbf{14},\mathbf{1})$ and $\Phi_R \sim (\mathbf{1}, \mathbf{1}, \mathbf{14})$, where $\Phi_L$ is real and $\Phi_R$ is complex, will do the job.  
The vev of $\Phi_L$ serves to `link together' the deconstructed $SU(2)_{L,i}$ factors, as was appreciated in Ref.~\cite{Kuo:1984md}. In terms of the $Sp(6)_L$ indices, we choose the vev to be
\be \label{eq:LprimeVEV}
\langle \Phi_L \rangle = \underbrace{\epsilon_L^{23} \Lambda_H \left(b_2\wedge b_6 + b_3\wedge b_5 \right)}_{\langle\phi_L^{23} \rangle} + \underbrace{\epsilon_L^{12} \Lambda_H \left(b_1 \wedge b_5 + b_2\wedge b_4\right)}_{\langle\phi_L^{12} \rangle}, 
\ee
where we have indicated how, in the $\Gint$-invariant intermediate theory, this vev ends up in two components that we denote by lower case symbols, in the representations
$\phi_L^{23} \sim [{\bf 1},({\bf 1},{\bf 2},{\bf 2}),{\bf 1}]_{{ 0}}$ and
$\phi_L^{12} \sim [{\bf 1},({\bf 2},{\bf 2},{\bf 1}),{\bf 1}]_{{ 0}}$.
The $\epsilon_L^{ij}$ are assumed to be small parameters (both $<1$), that each sets a separation of scales relative to the EFT matching scale $\Lambda_H$. The vev (\ref{eq:LprimeVEV}) breaks $\prod_{i=1}^3 SU(2)_{L,i} \to SU(2)_L$, giving two more $SU(2)_L$ triplets of $W^{\prime a}$ bosons.

The vev of the complex scalar field $\Phi_R$, written in terms of $Sp(6)_R$ indices, is
\begin{align}  \label{eq:RprimeVEV}
\langle \Phi_R \rangle = \underbrace{\Lambda_H \epsilon_{R}^{23}w_{23}c_2 \wedge c_6}_{\phi_{R}^{23}} + \underbrace{\Lambda_H \epsilon_{R}^{23}\overline{w}_{23} c_3 \wedge c_5}_{\overline{\phi}_{R}^{23}} + 
\underbrace{\Lambda_H\epsilon_{R}^{12} \left( w_{12}c_1 \wedge c_5 + \overline{w}_{12} c_2 \wedge c_4\right)}_{\phi_R^{12}}\, ,
\end{align}
where we take $w_{23}\overline{w}_{23}=1$ and $w_{12}\overline{w}_{12}=1$.
The $\Gint$ components that acquire the vev are in the representations $\phi_{R}^{23} \sim[{\bf 1},({\bf 1},{\bf 1},{\bf 1}),{\bf 4}]_{{+ 3}}$, $\overline{\phi}_{R}^{23} \sim[{\bf 1},({\bf 1},{\bf 1},{\bf 1}),{\bf 4}]_{{-3}}$, and $\phi_{R}^{12} \sim [{\bf 1},({\bf 1},{\bf 1},{\bf 1}),{\bf 5}]_{{0}}$.\footnote{For completeness, the scalar fields $\Phi_L$ and $\Phi_R$, which transform in the ${\bf 14}$-dimensional irreps of the UV $Sp(6)_L$ and $Sp(6)_R$ symmetries respectively, themselves decompose under the high scale symmetry breaking step $\GUV \to \Gint$ as
\begin{align}
\Phi_{L}\mapsto& [{\bf 1},({\bf 1},{\bf 1},{\bf 1}),{\bf 1}]_{0}^{\oplus 2} \oplus {[\underline{{\bf 1},({\bf 2},{\bf 2},{\bf 1}),{\bf 1}}]_{0}} \oplus [{\bf 1},({\bf 2},{\bf 1},{\bf 2}),{\bf 1}]_{0}\oplus {[\underline{{\bf 1},({\bf 1},{\bf 2},{\bf 2}),{\bf 1}}]_{0}}\, , \nonumber \\
\Phi_{R}\mapsto& [{\bf 1},({\bf 1},{\bf 1},{\bf 1}),{\bf 1}]_{0} \oplus [\underline{{\bf 1},({\bf 1},{\bf 1},{\bf 1}),{\bf 5}}]_{0} \oplus [\underline{{\bf 1},({\bf 1},{\bf 1},{\bf 1}),{\bf 4}}]_{-3}
 \oplus [\underline{{\bf 1},({\bf 1},{\bf 1},{\bf 1}),{\bf 4}}]_{3} \, . \label{eq:Phi_decomp} 
\end{align}
The underlined components are those that pick up vevs, as detailed in the main text.
}
Again, the parameters $\epsilon_{R}^{23}$ and $\epsilon_{R}^{12}$ encode ratios of scales with respect to $\Lambda_H$, and are assumed to be $<1$.
The vev (\ref{eq:RprimeVEV}) breaks $Sp(4)_{R,12}\times U(1)_R \to U(1)_Y$.
This gives rise to 10 heavy gauge bosons, decomposing as three (complex) $Z^\pm$ bosons in the representation $({\bf 1}, {\bf 1})_{6}$,
plus four neutral (real) $\ZP$ bosons. 

The end result of these breakings induced by $\Phi_L$ and $\Phi_R$ is the SM. To summarize, the scalar sector of the model is recorded in Table~\ref{tab:Scalars}, and we list all the 45 heavy gauge bosons that appear in our spectrum, together with the scales that set their masses, in Table~\ref{tab:heavyGB}.

\begin{table}
\begin{center}
\begin{tabular}{|c|c||c|c|}
\hline
& & Heavy scales ($\Lambda_{L,R}$) & Intermediate scale ($\epsilon \Lambda_H$) \\
\hline
Name & $\GSM$ representation & Number (origin) & Number (origin) \\ 
\hline
Charged $Z^\pm$ & $(\mathbf{1,1})_6$ & $3$ ($S_R$) & $3$ ($\Phi_R$)  \\
$U_1$ leptoquark & $(\mathbf{\overline{3},1})_{-4}$ & $1$ ($ S_R$) & $-$ \\
$W^{\prime a}$ triplet & $(\mathbf{1,3})_0\,(\mathbb{R})$ & $3$ ($ S_L$) &  $2$ ($\Phi_L$) \\
Real $\ZP$ & $(\mathbf{1,1})_0\, (\mathbb{R})$ &$3$ ($ S_L$), $5$ ($ S_R$) &  $4$ ($\Phi_R$) \\
 \hline
\end{tabular}
\end{center}
\caption{The decomposition of the 45 heavy gauge bosons in our theory, organised by the scales at which they obtain their mass. }
\label{tab:heavyGB}
\end{table}


\section{Quark masses and mixings} \label{sec:yukawa}

In this Section we show that realistic masses and mixings for the light quarks are naturally generated within the model we have set out, without any additional fields required. The idea is the following. When the heavy Higgs components are integrated out at the scale $M_{H}\approx \Lambda_H$, higher-dimensional Yukawa-like operators are generated in the $\Gint$-invariant intermediate EFT which involve insertions of the symmetry breaking fields $\Phi_L$ and $\Phi_R$. Once $\Phi_{L,R}$ acquire their vevs (\ref{eq:LprimeVEV}--\ref{eq:RprimeVEV}), which (a) link together different families due to their $Sp(6)_{L,R}$ orientation, and (b) are suppressed with respect to $M_{H}$ by small scale hierarchies (of order $10^{-1}$--$10^{-2}$), hierarchical Yukawa matrices are generated that have enough parametric freedom to account for the quark mass and mixing data. The built-in hierarchies mean the data are reproduced for `$\cO(1)$' couplings in the UV model.

By writing down a complete UV model and explicitly integrating out the relevant heavy degrees of freedom, we also ensure that no baryon number violating operators are generated in the IR that would trigger proton decay. One can straightforwardly see why this is the case, without going into details, by the following argument.
All terms present in our UV lagrangian will contract $SU(4)$ indices using one of the two inner products $\langle\cdot,\cdot\rangle_a$ defined in \S\ref{sec:notation} -- this is made explicit in our notation, {\em e.g.} in Eq.~(\ref{eq:V}) for the scalar potential. Hence, the EFT will not inherit any contractions of $SU(3)$ quark indices using the Levi-Civita tensor $\epsilon_{ijk}$. Since proton decay generically requires such contractions, it is absent from our model. As is the case in a renormalisable model with the Pati--Salam gauge group, but unlike in the $SU(5)$ model, a similar argument indicates the absence of proton decay mediated by the vector gauge bosons.

\subsection{EFT operators}

Before we dive into the details, it is helpful to first derive the fermion mass matrix hierarchies that we expect based on EFT reasoning alone. 
Using the EFT fields present in the $\Gint$-invariant theory, 
the leading Yukawa-like operators in the EFT expansion are
\begin{align}
\mathcal{L}\supset \sum_{a\in\{1,15\}}\begin{pmatrix}\frac{\phi_L^{23}\phi_L^{12}}{\Lambda_H^2}\overline{Q}_1 & \frac{\phi_L^{23}}{\Lambda_H}\overline{Q}_2 &\overline{Q}_3\end{pmatrix}\Bigg\{&\left(\overline{\mathcal{H}}_a {\mathcal{U}}_a^{1}+\mathcal{H}^\ast_a \overline{\mathcal{U}}_a^{1}\right)\frac{1}{\Lambda_H^2}
{\scriptstyle\begin{psmallmatrix}{\phi_R^{12}\phi_R^{23}}\\ {\phi_R^{12}\overline{\phi}_R^{23\ast}}\\{\phi_R^{12\ast}\phi_R^{23}}\\\phi_R^{12\ast}\overline{\phi}_R^{23\ast}\end{psmallmatrix}}\mathcal{Q}_{R,12}& \nonumber\\
&+\left(\overline{\mathcal{H}}_a {\mathcal{U}}_a^{2}+\mathcal{H}^\ast_a \overline{\mathcal{U}}_a^{2}\right)\frac{1}{\Lambda_H}{\scriptstyle\begin{psmallmatrix} \phi_R^{23} \\ \overline{\phi}_R^{23\ast}  \end{psmallmatrix}}\mathcal{Q}_{R,12}& \nonumber \\
&+\left(\overline{\mathcal{H}}_a {\mathcal{U}}_a^{3}
+\mathcal{H}^\ast_a \overline{\mathcal{U}}_a^{3}\right)U_3 \Bigg\}  \, .\label{eq:UYukOpsEFT}
\end{align}
This formula may appear complicated, and requires some explanation. Firstly, the three lines of the formula list operators that populate the three columns of the Yukawa matrix. The EFT coefficients are contained in matrices denoted 
$\mathcal{U}_a^i$, $i\in\{1,2,3\}$,
and their `barred' versions -- thus, the family index $i$ indicates the column number of the Yukawa matrix. Now, each of these coefficient matrices has 3 rows, which are contracted with the three columns of $\begin{pmatrix}\frac{\phi_L^{23}\phi_L^{12}}{\Lambda_H^2}\overline{Q}_1 & \frac{\phi_L^{23}}{\Lambda_H}\overline{Q}_2 &\overline{Q}_3\end{pmatrix}$. 
The columns of the coefficient matrices, on the other hand, are contracted with the column vectors containing combinations of $\phi_R$ fields that we have written explicitly. Thus, 
${\mathcal{U}}_a^{1}$ and $\overline{\mathcal{U}}_a^{1}$ are 3-by-4 matrices, 
${\mathcal{U}}_a^{2}$ and $\overline{\mathcal{U}}_a^{2}$ are 3-by-2 matrices, and 
${\mathcal{U}}_a^{3}$ and $\overline{\mathcal{U}}_a^{3}$ are 3-by-1 matrices.
Based only on the EFT, we would generically expect all these matrices to be populated by arbitrary $\cO (1)$ numbers.

Similarly for the down-type quarks we have 
\begin{align}
\mathcal{L}\supset \sum_{a\in\{1,15\}}\begin{pmatrix}\frac{\phi_L^{23}\phi_L^{12}}{\Lambda_H^2}\overline{Q}_1 & \frac{\phi_L^{23}}{\Lambda_H}\overline{Q}_2 &\overline{Q}_3\end{pmatrix}\Bigg\{&\left(\mathcal{H}_a {\mathcal{D}}_a^{1}+\overline{\mathcal{H}}^\ast_a \overline{\mathcal{D}}_a^{ 1}\right)\frac{1}{\Lambda_H^2}{\scriptstyle\begin{psmallmatrix}{\phi_R^{12}\overline{\phi}_R^{23}}\\ {\phi_R^{12}\phi_R^{23\ast}}\\{\phi_R^{12\ast}\overline{\phi}_R^{23}}\\\phi_R^{12\ast}{\phi}_R^{23\ast}\end{psmallmatrix}}\mathcal{Q}_{R,12}& \nonumber\\
&+\left(\mathcal{H}_a {\mathcal{D}}_a^{2}+\overline{\mathcal{H}}^\ast_a \overline{\mathcal{D}}_a^{2}\right)\frac{1}{\Lambda_H}{\scriptstyle\begin{psmallmatrix} \overline\phi_R^{23} \\ {\phi}_R^{23\ast}  \end{psmallmatrix}}\mathcal{Q}_{R,12}& \nonumber\\
&+\left(\mathcal{H}_a {\mathcal{D}}_a^{3}+\overline{\mathcal{H}}^\ast_a \overline{\mathcal{D}}_a^{3}\right)D_3 \Bigg\}\, ,  \label{eq:DYukOpsEFT}
\end{align}
for another set of {\em a priori} $\mathcal{O}(1)$ coefficient matrices ${\mathcal{D}}_a^{i}$ and their `barred' versions, and a very similar set of operators for the charged lepton Yukawas, 
\begin{align}
\mathcal{L}\supset \sum_{a\in\{1,15\}}\begin{pmatrix}\frac{\phi_L^{23}\phi_L^{12}}{\Lambda_H^2}\overline{L}_1 & \frac{\phi_L^{23}}{\Lambda_H}\overline{L}_2 &\overline{L}_3\end{pmatrix}\Bigg\{&\left(\mathcal{H}_a {\mathcal{E}}_a^{1}+\overline{\mathcal{H}}^\ast_a \overline{\mathcal{E}}_a^{1}\right)\frac{1}{\Lambda_H^2}{\scriptstyle\begin{psmallmatrix}{\phi_R^{12}\overline{\phi}_R^{23}}\\ {\phi_R^{12}\phi_R^{23\ast}}\\{\phi_R^{12\ast}\overline{\phi}_R^{23}}\\\phi_R^{12\ast}{\phi}_R^{23\ast}\end{psmallmatrix}}\mathscr{L}_{R,12}& \nonumber\\
&+\left(\mathcal{H}_a {\mathcal{E}}_a^{2}+\overline{\mathcal{H}}^\ast_a \overline{\mathcal{E}}_a^{2}\right)\frac{1}{\Lambda_H}{\scriptstyle\begin{psmallmatrix} \overline\phi_R^{23} \\ {\phi}_R^{23\ast}  \end{psmallmatrix}}\mathscr{L}_{R,12}& \nonumber\\
&+\left(\mathcal{H}_a {\mathcal{E}}_a^{3}+\overline{\mathcal{H}}^\ast_a \overline{\mathcal{E}}_a^{3}\right)E_3 \Bigg\} \, . \label{eq:EYukOpsEFT}
\end{align}
When the scalar fields $\Phi_{L,R}$ acquire their vevs at the lower scale, breaking $\Gint \to \GSM$ in the process, these operators match onto Yukawa couplings for all three families, with lighter family Yukawas coming from higher order operators in the EFT expansion.

To see this, it is first convenient to gather together combinations of the EFT coefficients, weighted by the factors $w_{ij}$ and $\overline{w}_{ij}$ that appear in the vev (\ref{eq:RprimeVEV}) of $\Phi_R$.
We thus define 3-by-3 matrices 
\begin{align} \label{eq:EFT_coeff_matrices}
{\mathcal{U}}_a&:=\left({\mathcal{U}}_a^{1} {\scriptstyle\begin{psmallmatrix}w_{12}w_{23} \\w_{12} \overline{w}_{23}^\ast\\ \overline{w}_{12}^\ast w_{23}\\  \overline{w}_{12}^\ast\overline{w}_{23}^\ast\end{psmallmatrix}},\, {\mathcal{U}}_a^{2}{\scriptstyle\begin{psmallmatrix} w_{23} \\ \overline{w}_{23}^\ast\end{psmallmatrix}}, \,  {\mathcal{U}}_a^{3}\right)\, , \nonumber\\ 
{\mathcal{D}}_a&:=\left({\mathcal{D}}_a^{1} {\scriptstyle\begin{psmallmatrix}\overline{w}_{12}\overline{w}_{23} \\ \overline{w}_{12} {w}_{23}^\ast\\ {w}_{12}^\ast \overline{w}_{23}\\  w_{12}^\ast {w}_{23}^\ast\end{psmallmatrix}}, \, {\mathcal{D}}_a^{2}{\scriptstyle\begin{psmallmatrix} \overline{w}_{23} \\ {w}_{23}^\ast\end{psmallmatrix}}, \, {\mathcal{D}}_a^{3}\right)\, , \nonumber \\
{\mathcal{E}}_a&:=\left({\mathcal{E}}_a^{1} {\scriptstyle\begin{psmallmatrix}\overline{w}_{12}\overline{w}_{23} \\ \overline{w}_{12} {w}_{23}^\ast\\ {w}_{12}^\ast \overline{w}_{23}\\  w_{12}^\ast {w}_{23}^\ast\end{psmallmatrix}}, \, {\mathcal{E}}_a^{2}{\scriptstyle\begin{psmallmatrix} \overline{w}_{23} \\ {w}_{23}^\ast\end{psmallmatrix}}, \, {\mathcal{E}}_a^{3}\right)\, , 
\end{align}
and define $\overline{\mathcal{U}}_a$, $\overline{\mathcal{D}}_a$, and $\overline{\mathcal{E}}_a$ analogously {\em i.e.} with the same structure in terms of $w_{ij}$ and $\overline{w}_{ij}$, just replacing each ${\mathcal{U}}^{i}_a$ by $\overline{\mathcal{U}}^{i}_a$ {\em etc}. Once the Higgs fields $\mathcal{H}_a$ and $\overline{\mathcal{H}}_a$ also acquire their vevs, we obtain the following mass matrices for the SM quarks:
\begin{align} 
\sqrt{2}M^u=\sum_{a\in\{1,15\}}\diag(\epsilon_L^{23}\epsilon_L^{12}, \epsilon_L^{23},1) (\overline{v}_a {\mathcal{U}}_a+v^\ast_a \overline{\mathcal{U}}_a)\diag(\epsilon_R^{23}\epsilon_R^{12}, \epsilon_R^{23},1)\, ,\label{eq:EFT_u-qu_mass} \\
\sqrt{2}M^d=\sum_{a\in\{1,15\}}\diag(\epsilon_L^{23}\epsilon_L^{12}, \epsilon_L^{23},1) ({v}_a {\mathcal{D}}_a+\overline{v}^\ast_a \overline{\mathcal{D}}_a)\diag(\epsilon_R^{23}\epsilon_R^{12}, \epsilon_R^{23},1)\, .  \label{eq:EFT_d-qu_mass}
\end{align}
The mass matrix $\sqrt{2}M^e$ for the charged leptons is given by a similar formula to $\sqrt{2}M^d$, except that the $a=15$ components are weighted in the sum by an overall factor of $-3$. Writing this out explicitly, we find all three SM fermion mass matrices have the following hierarchical structure
\begin{align} \label{eq:Mmatrix}
\frac{M^f}{v}
\sim \begin{pmatrix} 
\epsilon_L^{12}\epsilon_L^{23}\epsilon_R^{12}\epsilon_R^{23} &\epsilon_L^{12}\epsilon_L^{23}\epsilon_R^{23} &\epsilon_L^{12}\epsilon_L^{23}\\
\epsilon_L^{23}\epsilon_{R}^{12}\epsilon_{R}^{23} &\epsilon_L^{23}\epsilon_{R}^{23} &\epsilon_L^{23}\\
\epsilon_{R}^{12}\epsilon_{R}^{23} &\epsilon_{R}^{23} &1\\
\end{pmatrix}\, .
\end{align}
Note that in the limit where $\epsilon_L^{12} \to 1$ and $\epsilon_R^{12} \to 1$, the upper-left 2-by-2 block has the same suppression factor. Thus, the parameters $\epsilon_L^{12}$ and $\epsilon_R^{12}$ roughly act as spurions for $SU(2)$ symmetries acting on the first two generations of left- and right-handed fields respectively. Not surprisingly, the $\epsilon_L^{12}$ `spurion' will end up being of order the Cabibbo angle if we are to accurately model the observed quark masses and mixings in the 1-2 sector, as we see in \S \ref{sec:quark}.

\subsection{EFT matching formulae}

To see how these EFT operators are explicitly generated in our model without needing any additional fields, and to calculate the Wilson coefficients ${\mathcal{U}}_{1,15}$, ${\mathcal{D}}_{1,15}$, and ${\mathcal{E}}_{1,15}$, we must first discuss the scalar potential of the UV model. Since the fields $S_L$ and $S_R$, which recall trigger the first high scale symmetry breakings, are integrated out before the heavy components of the Higgs fields, we focus on the interactions between the Higgs fields $H_{1,15}$ and the symmetry breaking scalars $\Phi_{L,R}$. 

Interactions between $H_{1,15}$ and the $\Phi_{L,R}$ fields are governed by $\FSS$ gauge invariance, and there is a large number of independent terms that can be written down and so should be included. Before we do so, we think it important to clarify that, while quartic terms of the form $\sim H_a^4$ are of course required to generate the third-family aligned electroweak symmetry breaking vevs (\ref{eq:H1vev}--\ref{eq:H15vev}), we do not include them here because they do not affect the Yukawa-like operators that are generated upon integrating out the heavy Higgs components (to the order we are working).\footnote{As we observed in the Introduction, it is clear that the scalar potential of our model will need tuning in order to explain the particular vevs and masses of the scalars that we require. 
For now, we point out that the `alignment' of the Higgs vev with the third-family direction (which is first originates with the vev of $S_R$) does not require tuning. To justify this, consider the following interactions between $H_1$ and $S_{L/R}$:
\begin{equation}
M_H\langle H_1,H_1\rangle_1+\alpha_1\langle H_1, S_L \Omega H_1\rangle_1+\alpha_2\langle H_1,  H_1\Omega S_RS_R^\dagger  \rangle_1+\alpha_3\langle H_1,  H_1\Omega S_R^\dagger S_R  \rangle_1\, ,
\end{equation}
where $\alpha_i$ are free-parameters. Inserting the vevs of $S_L$ (\ref{eq:SLvev}) and $S_R$ (\ref{eq:SRvev}), and choosing order-1 parameter values $\alpha_1=-0.5 M_H, \alpha_{2}=-0.6M_H, \alpha_{3}=0.6M_H$ (say), the only non-zero negative eigenvalues of the resulting mass matrix sit in the direction of the third family. A similar argument can be applied to $H_{15}$. The inclusion of additional terms for large enough $M_H$ will not alter this fact.
A more detailed study of the potential and its requisite tuning will be explored in future studies.
} Similarly, we do not include quartic terms involving only $\Phi_L$ and $\Phi_R$.

Thus, for our purpose, a sufficient set of terms in the potential are the following interactions,\footnote{The potential here is the most general one quadratic in $H_a$ and invariant under a global $U(1)$-symmetry under which only the $H_a$ carry non-zero charges. Including additional terms only aids to increase the freedom to fit the data - in this sense, we consider this potential to be sufficient for our study.}
\begin{align} \label{eq:V}
V(H,\Phi) = \sum_{a\in \{1,15\}}\left( M_{H_a}^2\langle H_a,H_a\rangle_a-\langle H_a, f(H_a)\rangle_a\right) + \dots, 
\end{align}
where we have explicitly included the Higgs bare mass terms, and
where we found it convenient to define the operators $f_1:V_L \otimes V_R \to V_L \otimes V_R$ and $f_{15}:V_4 \otimes V_4^\ast \otimes V_L \otimes V_R \to V_4 \otimes V_4^\ast \otimes  V_L \otimes V_R$,
\begin{align}
f_a(A):=&-\Bigg\{\Lambda_H\beta_L^a \Phi_L \Omega A+\beta_{LL}^a \Phi_L \Omega \Phi_L \Omega A+\Lambda_H A\Omega \betaR{\scriptstyle\begin{psmallmatrix} \Phi_R\\ \Phi_R^\ast\end{psmallmatrix}} \nonumber \\
&+\Phi_L\Omega A \Omega \betaLR{\scriptstyle\begin{psmallmatrix} \Phi_R\\ \Phi_R^\ast\end{psmallmatrix}}
+A \Omega \betaRR {\scriptstyle\begin{psmallmatrix}\Phi_R\Omega\Phi_R \\ \Phi_R^\ast \Omega\Phi_R \\ \Phi_R\Omega \Phi_R^\ast \\ \Phi_R^\ast\Omega \Phi_R^\ast \end{psmallmatrix}}\Bigg\}\, .
\end{align}
These $f_a(A)$ are mass dimension-3 operators, one for each value of $a \in \{1,15\}$, which encode all the cubic and quartic interactions between $H_a$ and $\Phi_{L,R}$.
Here $\beta_L^a$ and $\beta_{LL}^a$ are dimensionless real coupling constants, and the following are vectors of dimensionless coupling constants:
\begin{align}
\betaR&:=\left(\beta_{R}^a, \beta_{R}^{a\ast}\right) \text{ with } \beta_{R}^a\in \mathbb{C},\label{eq:betaR_def}\\
\betaLR&:=\left(\beta_{LR}^a, \beta_{LR}^{a\ast}\right) \text{ with } \beta_{LR}^a\in \mathbb{C},\label{eq:betaLR_def}\\
\betaRR&:= \left(\beta^a_{RR}, \tilde \beta^a_{RR}, \check \beta^a_{RR} ,\beta_{RR}^{a\ast}\right)\text{ with }  \beta^a_{RR}\in \mathbb{C},\;  \tilde \beta^a_{RR}, \check \beta^a_{RR}\in \mathbb{R}\label{eq:betaRR_def}\, .
\end{align}
We observe  that the operators $f_a(A)$ are hermitian with respect to both inner products $\langle\cdot,\cdot\rangle_a$ defined in (\ref{eq:innerP_1}) and (\ref{eq:innerP_15}), namely
$\langle A,f_a(B)\rangle_a=\langle f_a(A),B\rangle_a$
for all $A$ and $B$ (in the appropriate domain). 

With such vertices, one can write down Feynman diagrams in the UV model that link fermions with different family indices to the Higgs components $\mathcal{H}_{1,15}$ and $\overline{\mathcal{H}}_{1,15}$ that acquire the EWSB vevs. The $\Phi_{L,R}$ fields, which recall transform in 2-index ${\bf 14}$-dimensional irreps of $Sp(6)_{L,R}$, play a crucial role in the flavour structure of these vertices; their 2-index vevs can be thought of as matrices that `transfer' the 3$^{\text{rd}}$ family-aligned Higgs vev (in the $b_3 \otimes c_6$ and $b_6 \otimes c_3$ directions) to the light family $Sp(6)_{L,R}$ indices $b_{1, 2, 4, 5}$ and $c_{1, 2, 4, 5}$. These indices are carried by components of the {\em heavy} Higgs fields that are integrated out, which then couple to the light family fermions via the same Yukawa couplings (\ref{eq:ren_yuk}) that we wrote down in the renormalisable model.

In the next few Subsections we give the gory details of the EFT matching procedure, which appears complicated in large part due to the fact that there are many different interactions in the scalar sector that are similar but have independent couplings. This feature, which follows generically given our scalar field content, is in fact important in giving enough freedom in the resulting EFT coefficients to fully explain the CKM matrix, and the differences between the up and down quark and charged lepton spectra. For more casual readers, we recommend skipping ahead to Figs.~\ref{fig:dim5}--\ref{fig:11} for the relevant Feynman diagrams, which conveniently summarize how all the effective Yukawa couplings are generated.

\subsubsection*{Integrating out the heavy Higgses}

It is easy enough, albeit a little tedious, to explicitly integrate out the heavy Higgs components at tree level.
Firstly, let us define a 3rd-family projection map $P_3$, which projects each $H_a$ onto its vev-acquiring irreps of $\Gint$ (we emphasize that these are the {\em light} degrees of freedom in $H_{1,15}$). We also define $P_{12}:=\mathrm{id}-P_3$. 
The projections $P_3$ and $P_{12}$ project $H_a$ (and other elements in the appropriate vector space) onto subspaces that are orthogonal under our inner products, namely
\begin{align}
\langle P_3 A, P_{12}B\rangle_a=\langle P_{12}A,P_3 B\rangle_a=0
\end{align}
for any $A$ and $B$.

We intend to integrate out the heavy Higgs components $P_{12} H_a$ and $P_{12} H_a^\ast$ at tree-level, and for that we need their equations of motion.
We have the following general formulae for functional derivatives of our inner products with respect to the heavy Higgs fields,
\begin{align}
\frac{D}{DP_{12}H_a}\int \langle A, H_a\rangle_a \, d^4x=\Omega^TP_{12}A^\ast\Omega,\quad
\frac{D}{DP_{12}H^\ast_a}\int \langle H_a,A\rangle_a \, d^4x=\Omega P_{12}A\Omega^T,
\end{align}
which we can use to differentiate the Yukawa couplings~(\ref{eq:ren_yuk}) and the terms in the potential that we wrote explicitly in (\ref{eq:V}).
Solving the equations of motion for $P_{12} H$, we get 
\begin{align}
P_{12}H_a=\frac{1}{M_H^2} \overline{y}_aP_{12} (\Gamma_a\Psi_R\overline{\Psi}_L)^T+\frac{1}{M_H^2} {y}_a^\ast P_{12} (\Gamma_a\Psi_R\overline{\Psi}_L)^\dagger+\frac{1}{M_H^2} P_{12}f_a(H_a) + \dots, 
\end{align}
where the `$+\dots$' indicates contributions from all the additional terms in the potential that we are not writing explicitly (such as the various $H_a^4$ terms).
This admits a series solution
\begin{align} \label{eq:P12_EOM}
P_{12}H_a=&\sum_{m=0}^\infty \frac{1}{M_H^{2(m+1)}}(P_{12}f_{a})^m(P_{12}(\overline{y}_a(\Gamma_a\Psi_R\overline{\Psi}_L)^T + y_a^\ast (\Gamma_a\Psi_R\overline{\Psi}_L)^\dagger) \nonumber \\
+&\sum_{n=1}^\infty \frac{1}{M_H^{2n}}(P_{12}f)^n(P_{3}H_a) +\dots
\end{align}
To get the effective action that follows from integrating out $P_{12} H_a$, we substitute (\ref{eq:P12_EOM}) back into the action.
The terms in the effective action that we are interested in are the Yukawa-type ones.  Substituting (\ref{eq:P12_EOM}) into the renormalisable Yukawa terms (\ref{eq:ren_yuk}), we obtain the following towers of EFT operators
\begin{align}
\cL_{\mathrm{yuk}} &\supset y_a \langle (\Gamma_a\Psi_{R}\overline{\Psi}_{L} )^\dagger,H_a\rangle_{a}&\mapsto \sum_{n=0}^\infty \frac{y_a
}{M_H^{2n}} \langle (\Gamma_a\Psi_{R}\overline{\Psi}_{L} )^\dagger, (P_{12}f_a)^n (P_3H_a)\rangle_{a}, \nonumber\\
\cL_{\mathrm{yuk}} &\supset \overline{y}_a\langle H_a, (\Gamma_a\Psi_{R}\overline{\Psi}_{L} )^T\rangle_{a}&\mapsto \sum_{n=0}^\infty \frac{\overline{y}_a}{M_H^{2n}}\langle (P_{12}f_a)^n (P_3H_a),(\Gamma_a\Psi_{R}\overline{\Psi}_{L} )^T\rangle_a, \label{eq:EFT_yuk_tower}
\end{align}
plus equivalent formulae for their conjugates.
We checked that substituting (\ref{eq:P12_EOM}) into the potential itself gives a vanishing tree-level contribution to Yukawa operators.

\subsubsection*{Detailed example: dimension 5 Yukawa operators}

There are four dimension-5 terms in the EFT expansion of $\cL_{\mathrm{yuk}}$, encoded in (\ref{eq:EFT_yuk_tower}). Two are given by 
\begin{align}
\cL_{\mathrm{yuk}} \supset -
\sum_{a\in\{1,15\}}\bigg\{&\frac{y_a\Lambda_H}{M_H^{2}}\langle(\Gamma_a\Psi_{R}\overline{\Psi}_{L} )^\dagger,P_{12}(\beta_L \Phi_L \Omega P_3 H_a)\rangle_a \nonumber \\
&+\frac{\overline{y}_a\Lambda_H}{M_H^{2}}\langle P_{12}(\beta_L \Phi_L \Omega P_3 H_a),(\Gamma_a\Psi_{R}\overline{\Psi}_{L} )^T\rangle_a\bigg\}\, ,
\end{align}
where we have used the fact that the projection $P_{12}$ commutes with the transpose. 
Once $\Phi_L$ is expanded around its vev (\ref{eq:LprimeVEV}), which recall breaks the deconstructed weak symmetry to its diagonal subgroup, $\langle \Phi_L \rangle: \prod SU(2)_{L,i} \to SU(2)_L$, these dimension-5 terms match onto dimension-4 Yukawa couplings of the SM with an EFT suppression factor of $\epsilon_L^{23}$. 
Further expanding the Higgs about its vev, the resulting mass terms are
\begin{multline} \label{eq:2-3-mass}
\frac{\Lambda_H^2}{2M_H^2} \epsilon_L^{23}\big\{\beta_L^1(y_1 v_1+\overline{y}_1 \overline v_1^\ast)(\overline{D}_{L2}D_{R3}+\overline{E}_{L2} E_{R3})+\beta_L^{15}(y_{15} v_{15}+\overline{y}_{15} \overline v_{15}^\ast)(\overline{D}_{L2}D_{R3}-3\overline{E}_{L2} E_{R3})\\
+\beta_L^1(y_1 \overline{v}_1+\overline{y}_1 v_1^\ast)( \overline{U}_{L2}U_{R3}+\overline{\nu}_{2L}\nu_{3R})+\beta_L^{15}(y_{15} \overline v_{15}+\overline{y}_{15} v_{15}^\ast)( \overline{U}_{L2}U_{R3}-3\overline{\nu}_{2L}\nu_{3R})\big\}\, .
\end{multline}
Assuming hereon that 
$M_H$ equals $\Lambda_H$ the EFT matching scale,
for simplicity,
we can thus read off the following EFT coefficients,
\begin{align}
[\,\mathcal{D}_a^{3}\,]_{2}&=[\,\mathcal{U}_a^{3}\,]_{2}=\frac{1}{2}y_a\beta_L^a, \quad  [\,\mathcal{E}_1^{3}\,]_{2}=\frac{1}{2}y_1\beta_L^1,\quad  [\,\mathcal{E}_{15}^{3}\,]_{2}=-\frac{3}{2}y_{15}\beta_L^{15},\\
[\,\overline{\mathcal{D}}_a^{3}\,]_{2}&=[\,\overline{\mathcal{U}}_a^{3}\,]_{2}=\frac{1}{2}\overline y_a\beta_L^a, \quad  [\,\overline{\mathcal{E}}_1^{3}\,]_{2}=\frac{1}{2}\overline y_1\beta_L^1,\quad  [\,\overline{\mathcal{E}}_{15}^{3}\,]_{2}=-\frac{3}{2}\overline y_{15}\beta_L^{15},
\end{align}
which populate the 2-3 elements of each fermion mass matrix. Note the relative factors of $-3$ appearing in the $SU(4)$-adjoint Higgs' couplings to charged leptons.

The other contributions from dimension-5 terms in $\mathcal{L}_Y$ are given by 
\begin{align}
\cL_{\mathrm{yuk}} \supset 
-\sum_{a\in\{1,15\}}\bigg\{&\frac{y_1\Lambda_H}{M_H^{2}}\langle(\Gamma_a\Psi_{R}\overline{\Psi}_{L} )^\dagger,P_{12}( P_3 H_a\Omega \betaR {\scriptstyle\begin{psmallmatrix} \Phi_R\\ \Phi_R^\ast\end{psmallmatrix}})\rangle_a \nonumber \\
&+\frac{\overline{y}_1\Lambda_H}{M_H^{2}}\langle P_{12}( P_3 H_a\Omega \betaR {\scriptstyle\begin{psmallmatrix} \Phi_R\\ \Phi_R^\ast\end{psmallmatrix}}),(\Gamma_a\Psi_{R}\overline{\Psi}_{L} )^T\rangle_a\bigg\}\, ,
\end{align}
and these populate the 3-2 elements of the fermion mass matrices.
Instead of writing out the full expressions for these mass terms as per (\ref{eq:2-3-mass}), we are here content to just write down the contributions to  the relevant EFT coefficients. These are
\begin{align}
[\,\mathcal{D}_a^{2}\,]_{3\star}&=[\,\mathcal{U}_a^{2}\,]_{3\star}=\frac{1}{2}y_a\betaR, \quad  
[\,\mathcal{E}_1^{2}\,]_{3\star}=\frac{1}{2}y_1\boldsymbol{\beta}_{R}^1,\quad 
[\,\mathcal{E}_{15}^{2}\,]_{3\star}=-\frac{3}{2}y_{15}\boldsymbol{\beta}_{R}^{15}, \\
[\,\overline{\mathcal{D}}_a^{2}\,]_{3\star}&=[\,\overline{\mathcal{U}}_a^{2}\,]_{3\star}=\frac{1}{2}\overline y_a\betaR,\quad 
[\,\overline{\mathcal{E}}_1^{2}\,]_{3\star}=\frac{1}{2}\overline y_1\boldsymbol{\beta}_{R}^1,\quad  
[\,\overline{\mathcal{E}}_{15}^{2}\,]_{3\star}=-\frac{3}{2}\overline y_{15}\boldsymbol{\beta}_{R}^{15},
\end{align}
where the $\star$ is a standard short-hand denoting the matrix rows. (Recall that the bold-face ${\boldsymbol{\beta}}$'s are themselves row-vectors, as defined in (\ref{eq:betaR_def}--\ref{eq:betaRR_def})).

\begin{figure}[htbp]
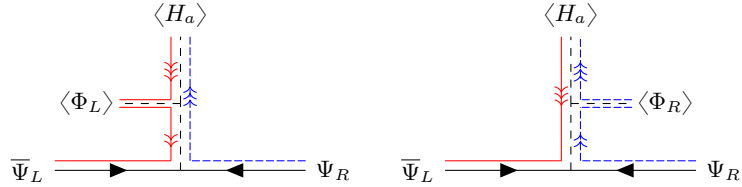

\begin{center}
\colordiagram{
    \fermions
    \HiggsII{L3}{R1}
    \phiLI{c1}
    \LII{L0}{L1}
    \LIII{L2}{L3}
    \RIII{R0}{R1}
}
\colordiagram{
    \fermions
    \HiggsII{L1}{R3}
    \phiRI{c1}
    \LIII{L0}{L1}
    \RII{R0}{R1}
    \RIII{R2}{R3}
}
\end{center}
\caption{Feynman diagrams that contribute to the 2-3 (left) and 3-2  (right) elements of the Yukawa matrices, once the heavy Higgs components running along the internal lines are integrated out at $\Lambda_H = M_H$. \label{fig:dim5}}
\end{figure}

The contributions that we have just derived to the 2-3 and 3-2 elements of the Yukawa matrices can be vizualized in terms of the tree-level Feynman diagrams in Fig.~\ref{fig:dim5}.

\noindent In what follows, for the higher dimensional operators, we will simply show the corresponding Feynman diagrams, and the contributions to the $\mathcal{D}_a$, $\mathcal{U}_a$, and $\mathcal{E}_a$ coefficient matrices (and their barred versions). Our life is made a little easier by the fact that some simple relations exist between the different matrices, namely
\begin{align} \label{eq:h-relations}
\mathcal{D}^{i}_a={\mathcal{U}}^{i}_a,\quad \overline{\mathcal{D}}^{i}_a=\overline{\mathcal{U}}^{i}_a,\quad {\mathcal{D}}^{i}_a=\left(\frac{\overline{y}_a}{y_a}\overline{\mathcal{D}}^{i}_a\text{ with } \tilde \beta_{RR}^a\leftrightarrow \check \beta_{RR}^a\right), \nonumber \\
{\mathcal{E}}^{i}_1={\mathcal{D}}^{i}_1,\quad {\mathcal{E}}^{i}_{15}=-3{\mathcal{D}}^{i}_{15},\quad 
\overline{\mathcal{E}}^{i}_1=\overline{\mathcal{D}}^{i}_1,\quad 
\overline{\mathcal{E}}^{i}_{15}=-3\overline{\mathcal{D}}^{i}_{15}\, .
\end{align}
Thus in what follows, we only explicitly write the contributions to the matrices $\mathbf{h}_a^d$, and all others can be inferred using (\ref{eq:h-relations}).

\subsubsection*{Dimension 6, 7, and 8 Yukawa operators}

\begin{figure}[htbp]
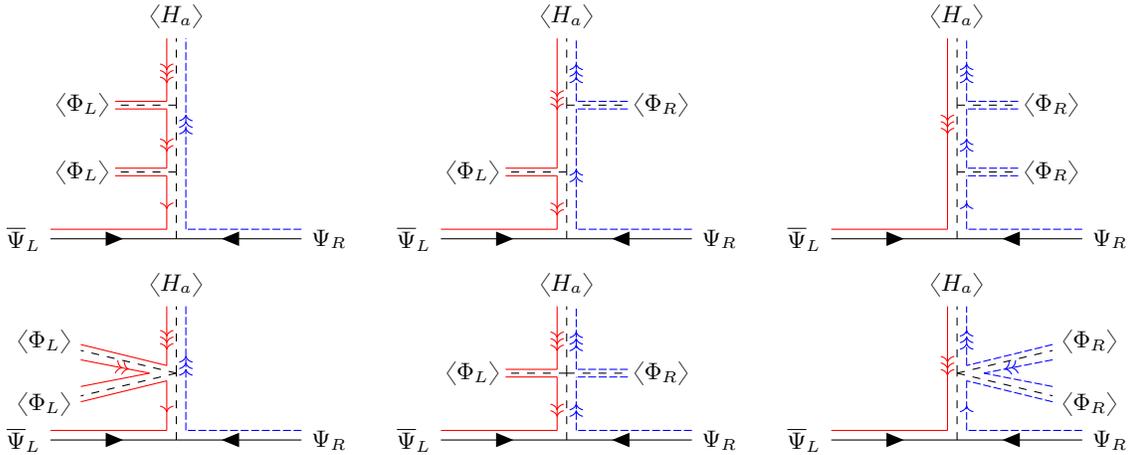

\begin{center}
\colordiagram{
    \fermions
    \HiggsIII{L5}{R1}
    \phiLI{c1}
    \phiLII{c2}
    \LI{L0}{L1}
    \LII{L2}{L3}
    \LIII{L4}{L5}
    \RIII{R0}{R1}
}
\colordiagram{
    \fermions
    \HiggsIII{L3}{R3}
    \phiRI{c2}
    \phiLI{c1}
    \RII{R0}{R1}
    \RIII{R2}{R3}
    \LII{L0}{L1}
    \LIII{L2}{L3}
}
\colordiagram{
    \fermions
    \HiggsIII{L1}{R5}
    \phiRI{c1}
    \phiRII{c2}
    \RI{R0}{R1}
    \RII{R2}{R3}
    \RIII{R4}{R5}
    \LIII{L0}{L1}
}
\colordiagram{
    \fermions
    \HiggsII{L3}{R1}
    \phiLL{c1}
    \LI{L0}{L1} 
    \LIII{L2}{L3}
    \RIII{R0}{R1}
}
\colordiagram{
    \fermions
    \HiggsII{L3}{R3}
    \phiRI{c1}
    \phiLI{c1}
    \RII{R0}{R1}
    \RIII{R2}{R3}
    \LII{L0}{L1}
    \LIII{L2}{L3}
}
\colordiagram{
    \fermions
    \HiggsII{L1}{R3}
    \phiRR{c1}
    \RI{R0}{R1} 
    \RIII{R2}{R3}
    \LIII{L0}{L1}
}
\end{center}
\caption{Feynman diagrams contributing to the 1-3 (left), 2-2 (middle) and 3-1 (right) elements of the Yukawa matrices. \label{fig:dim_6}}
\end{figure}

\begin{figure}[H]
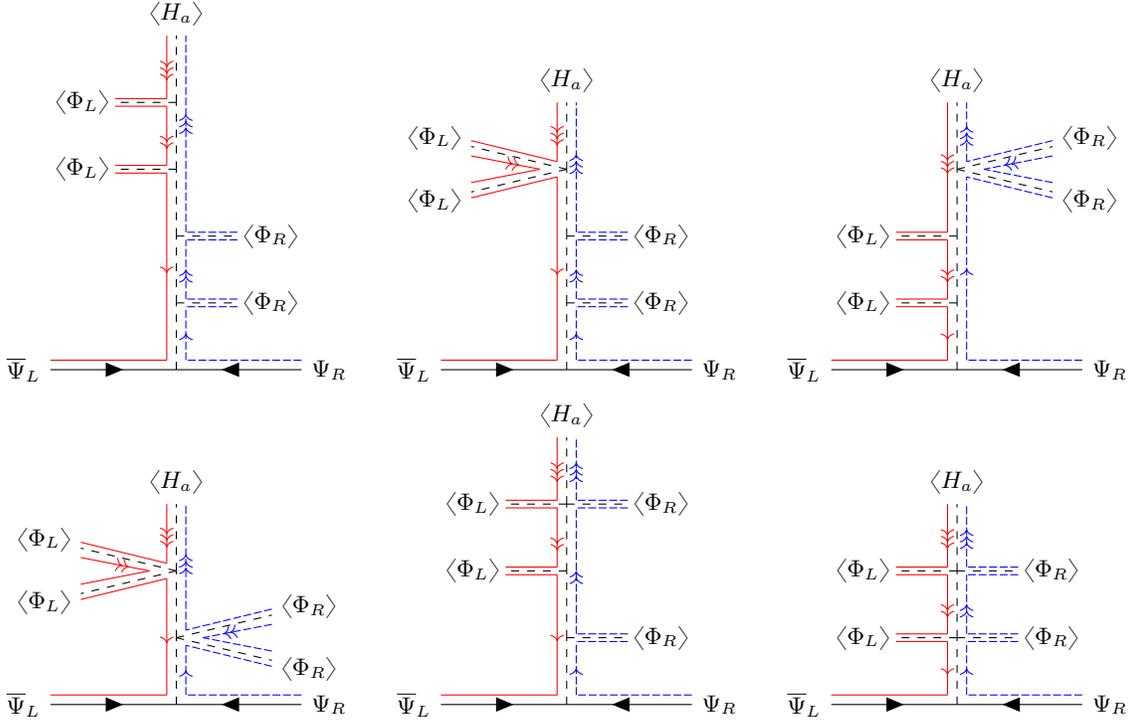

\begin{center}
\colordiagram{
    \fermions
    \HiggsV{L5}{R5}
    \phiLI{c3}
    \phiLII{c4}
    \phiRI{c1}
    \phiRII{c2}
    \LI{L0}{L1}
    \LII{L2}{L3}
    \LIII{L4}{L5}
    \RI{R0}{R1}
    \RII{R2}{R3}
    \RIII{R4}{R5}
}
\colordiagram{
    \fermions
    \HiggsIV{L3}{R5}
    \phiLL{c3}
    \phiRI{c1}
    \phiRII{c2}
    \LI{L0}{L1}
    \LIII{L2}{L3}
    \RI{R0}{R1}
    \RII{R2}{R3}
    \RIII{R4}{R5}
}
\colordiagram{
    \fermions
    \HiggsIV{L5}{R3}
    \phiLI{c1}
    \phiLII{c2}
    \phiRR{c3}
    \LI{L0}{L1}
    \LII{L2}{L3}
    \LIII{L4}{L5}
    \RI{R0}{R1}
    \RIII{R2}{R3}
}
\colordiagram{
    \fermions
    \HiggsIII{L3}{R3}
    \phiLL{c2}
    \phiRR{c1}
    \LI{L0}{L1}
    \LIII{L2}{L3}
    \RI{R0}{R1}
    \RIII{R2}{R3}
}
\colordiagram{
    \fermions
    \HiggsIV{L5}{R5}
    \phiLI{c2}
    \phiLII{c3}
    \phiRI{c1}
    \phiRII{c3}
    \LI{L0}{L1}
    \LII{L2}{L3}
    \LIII{L4}{L5}
    \RI{R0}{R1}
    \RII{R2}{R3}
    \RIII{R4}{R5}
}
\colordiagram{
    \fermions
    \HiggsIII{L5}{R5}
    \phiLI{c1}
    \phiLII{c2}
    \phiRI{c1}
    \phiRII{c2}
    \LI{L0}{L1}
    \LII{L2}{L3}
    \LIII{L4}{L5}
    \RI{R0}{R1}
    \RII{R2}{R3}
    \RIII{R4}{R5}
}
\end{center}
\caption{Feynman diagrams for the 1-1 elements of the Yukawa matrices.  \label{fig:11}}
\end{figure}

The dimension-6 operators contribute to the 1-3, 2-2 and 3-1 Yukawa couplings, {\em i.e.} those on the $i+j=4$ diagonal. The EFT coefficients are
\begin{align}
[\,\mathcal{D}^{3}_a\,]_{1\star}&=\frac{1}{2^2}y_a((\beta_L^a)^2-\beta_{LL}^a), \nonumber \\
[\,\mathcal{D}^{2}_a\,]_{2\star}&=\frac{1}{2^2}y_a(2\beta_L^a\betaR-\betaLR),\nonumber \\
[\,\mathcal{D}^{1}_a\,]_{3\star}&=\frac{1}{2^2} y_a(\betaRbetaR-\betaRR), 
\end{align}
corresponding to the Feynman diagrams in Fig.~\ref{fig:dim_6}.
The dimension-7 operators contribute to the 1-2 and 2-1 Yukawa couplings, {\em i.e.} those on the $i+j=3$ diagonal. These give
\begin{align}
[\,\mathcal{D}^{2}_a\,]_{1\star}&=\frac{1}{2^3}y_a(3\betaR (\beta_L^1)^2-2 \betaR \beta_{LL}^1-2\betaLR \beta_L^1),\\
[\,\mathcal{D}^{1}_a\,]_{2\star}&=\frac{1}{2^3}y_a(3 \beta_L^1\betaRbetaR-2\beta_L^1 \betaRR-\betaLR\otimes\betaR-\betaR\otimes\betaLR),
\end{align}
corresponding to the diagrams in Fig.~\ref{fig:dim_7}.
Lastly, the dimension-8 terms contribute to the 1-1 Yukawa coupling. These give
\begin{align}
[\,\mathcal{D}^{1}_a\,]_{1\star}&=\frac{1}{2 ^4}y_a (6(\beta_L^a)^2\betaRbetaR-3\beta_{LL}\betaRbetaR
-3(\beta_L^a)^2\betaRR+2\beta_{LL}\betaRR
\nonumber\\&
-3\beta_L\betaLR\otimes \betaR-3\beta_L\betaR\otimes\betaLR
+\betaLR\otimes\betaLR),
\end{align}
corresponding to the diagrams in Fig.~\ref{fig:11}.

Using all these formulae for the EFT coefficients, we can piece together the mass matrices using Eqs.~(\ref{eq:EFT_coeff_matrices}--\ref{eq:EFT_d-qu_mass}).

\begin{figure}[H]
\begin{center}
\colordiagram{
    \fermions
    \HiggsIV{L5}{R3}
    \phiRI{c3}
    \phiLI{c1}
    \phiLII{c2}
    \RII{R0}{R1}
    \RIII{R2}{R3}
    \LI{L0}{L1}
    \LII{L2}{L3}
    \LIII{L4}{L5}
}
\colordiagram{
    \fermions
    \HiggsIII{L3}{R3}
    \phiRI{c1}
    \phiLL{c2}
    \RII{R0}{R1}
    \RIII{R2}{R3}
    \LI{L0}{L1}
    \LIII{L2}{L3}
}
\colordiagram{
    \fermions
    \HiggsIII{L5}{R3}
    \phiRI{c2}
    \phiLI{c1}
    \phiLII{c2}
    \RII{R0}{R1}
    \RIII{R2}{R3}
    \LI{L0}{L1}
    \LII{L2}{L3}
    \LIII{L4}{L5}
}

\colordiagram{
    \fermions
    \HiggsIV{L3}{R5}
    \phiLI{c3}
    \phiRI{c1}
    \phiRII{c2}
    \LII{L0}{L1}
    \LIII{L2}{L3}
    \RI{R0}{R1}
    \RII{R2}{R3}
    \RIII{R4}{R5}
}
\colordiagram{
	\fermions
    \HiggsIII{L3}{R3}
    \phiLI{c1}
    \phiRR{c2}
    \LII{L0}{L1}
    \LIII{L2}{L3}
    \RI{R0}{R1}
    \RIII{R2}{R3}
}
\colordiagram{
	\fermions
    \HiggsIII{L3}{R5}
    \phiLI{c2}
    \phiRI{c1}
    \phiRII{c2}
    \LII{L0}{L1}
    \LIII{L2}{L3}
    \RI{R0}{R1}
    \RII{R2}{R3}
    \RIII{R4}{R5}
}
\end{center}
\caption{Feynman diagrams for the 2-1 (top) and 1-2 (bottom) elements of the Yukawa matrices. \label{fig:dim_7}}
\end{figure}

\subsection{Perturbative analysis of quark masses and mixings} \label{sec:quark}

We now derive formulae for the quark and charged lepton masses and the quark mixing angles predicted by the model. We begin by defining shorthand matrices of dimensionless coefficients
\begin{align}
&{\mathcal{U}} :=\frac{\sqrt{2}}{v}\sum_a\left( \overline{v}_{a}\,{\mathcal{U}}_{a} + v_{a}^\ast \,\overline{\mathcal{U}}_{a} \right) \, , \\
&{\mathcal{D}} :=\frac{\sqrt{2}}{v}\sum_a\left( v_{a}\,{\mathcal{D}}_{a} + \overline{v}_{a}^\ast\,\overline{\mathcal{D}}_{a} \right) \, , \\
&{\mathcal{E}} :=\frac{\sqrt{2}}{v}\sum_a\left( v_{a}\,{\mathcal{E}}_{a} + \overline{v}_{a}^\ast\,\overline{\mathcal{E}}_{a} \right) \, ,
\end{align}
where the Wilson coefficients ${\mathcal{U}}_{a}$, ${\mathcal{D}}_{a}$, ${\mathcal{E}}_{a}$, $\overline{\mathcal{U}}_{a}$, $\overline{\mathcal{D}}_{a}$, and $\overline{\mathcal{E}}_{a}$ can be written in terms of the fundamental couplings of the UV model using the formulae in the preceding Subsections. 

\subsubsection*{Leading order mass formulae}

Assuming all four parameters $\epsilon_{L(R)}^{ij}$ are $\ll 1$, we can use matrix perturbation theory to write down the leading expressions for the quark masses and mixing angles. For each of the matrices $\mathcal{U}$, $\mathcal{D}$, and $\mathcal{E}$, let $\widehat{\mathcal{U}}_{ij}$, $\widehat{\mathcal{D}}_{ij}$, and $\widehat{\mathcal{E}}_{ij}$ denote the minor obtained by removing the $i^\text{th}$ row and $j^\text{th}$ column of the respective matrix, then taking the determinant. For the masses we find, {for $f_1 \in \{u,d,e\}$, $f_2 \in \{c,s,\mu\}$, $f_3 \in \{t,b,\tau\}$, and corresponding $\mathcal{F} \in \{\mathcal{U}, \mathcal{D}, \mathcal{E}\}$, }
\begin{align}\label{eq:singular_values}
y_{f_1} &\approx \left|\frac{\det({\mathcal{F}})}{{\widehat{\mathcal{F}}}_{11}}\right| \epsilon_L^{12} \epsilon_{R}^{12}\epsilon_L^{23}\epsilon_{R}^{23}, \\
y_{f_2} &\approx\left|  \frac{{\widehat{\mathcal{F}}}_{11}}{{\mathcal{F}}_{33}} \right|\epsilon_L^{23}\epsilon_{R}^{23}, \\
y_{f_3} &\approx\left| {\mathcal{F}}_{33}\right|\, , 
\end{align}
where $y_{f_i} = \sqrt{2} m_{f_i}/v$ for each particle. The last equation, for the third family fermion masses, simply matches (\ref{eq:3rd-fam-masses}). Thus, the mass hierarchies between families are set by the expansion parameters $\epsilon_{L,R}^{ij}$. We emphasize that one could substitute in expressions for the EFT Wilson coefficients in terms of the fundamental UV couplings, but, with the exception of the renormalisable third family masses, the resulting formulae would be very complicated -- and not especially enlightening.

\subsubsection*{Fixing the scale separations}

We now turn to the mixing angles. Again using perturbation theory, we can find the unitary matrices $V^u_L$ and $V^u_R$ such that $V_L^u M^u(V_L^u M^u)^\dagger$ and $(M^uV_R^u)^\dagger M^uV_R^u$ are diagonal, and similarly for $V^d_L$ and $V^d_R$. The CKM matrix is then $\CKM = V^u_L V^{d\, \dagger}_L$. In terms of the small parameters, the result is
\begin{align} \label{eq:I}
&
 \CKM \approx 
 \end{align}
 \vspace{-0.65in}
 \begin{widepageII}
 \begin{align}
&{\small\begin{pmatrix}
1 -\left(\left|\frac{\widehat{\mathcal{D}}_{21}}{\widehat{\mathcal{D}}_{11}}\right|^2+\left|\frac{\widehat{\mathcal{U}}_{21}}{\widehat{\mathcal{U}}_{11}}\right|^2-2\frac{\widehat{\mathcal{U}}_{21}^\ast\widehat{\mathcal{D}}_{21}}{\widehat{\mathcal{U}}_{11}^\ast\widehat{\mathcal{D}}_{11}}\right)\frac{(\epsilon_L^{12})^2}{2}& \left( \frac{\widehat{\mathcal{D}}_{21}}{\widehat{\mathcal{D}}_{11}}-\frac{\widehat{\mathcal{U}}_{21}^\ast}{\widehat{\mathcal{U}}_{11}^\ast}\right)\epsilon_L^{12} & \left(\frac{\widehat{\mathcal{U}}_{31}^{
\ast}}{\widehat{\mathcal{U}}_{11}^\ast}+\frac{\mathcal{D}_{13}}{\mathcal{D}_{33}}-\frac{\mathcal{D}_{23}}{\mathcal{D}_{33}}\frac{\widehat{\mathcal{U}}_{21}^\ast}{\widehat{\mathcal{U}}_{11}^\ast}\right)\epsilon_L^{12} \epsilon_L ^{23}\\
\left( \frac{\widehat{\mathcal{U}}_{21}}{\widehat{\mathcal{U}}_{11}}-\frac{\widehat{\mathcal{D}}_{21}^\ast}{\widehat{\mathcal{D}}_{11}^\ast}\right)\epsilon_L^{12} & 1 -\left(\left|\frac{\widehat{\mathcal{D}}_{21}}{\widehat{\mathcal{D}}_{11}}\right|^2+\left|\frac{\widehat{\mathcal{U}}_{21}}{\widehat{\mathcal{U}}_{11}}\right|^2-2\frac{\widehat{\mathcal{U}}_{21}\widehat{\mathcal{D}}_{21}^\ast}{\widehat{\mathcal{U}}_{11}\widehat{\mathcal{D}}_{11}^\ast}\right)\frac{(\epsilon_L^{12})^2}{2}& \left(\frac{\mathcal{D}_{23}}{\mathcal{D}_{33}}-\frac{\mathcal{U}_{23}^\ast}{\mathcal{U}_{33}^\ast}\right)\epsilon_L^{23} \\
 \left(\frac{\widehat{\mathcal{D}}_{31}^\ast}{\widehat{\mathcal{D}}_{11}^\ast}+\frac{\mathcal{U}_{13}}{\mathcal{U}_{33}}-\frac{\mathcal{U}_{23}}{\mathcal{U}_{33}}\frac{\widehat{\mathcal{D}}_{21}^\ast}{\widehat{\mathcal{D}}_{11}^\ast}\right)\epsilon_L^{12} \epsilon_L^{23} & \left(\frac{\mathcal{U}_{23}}{\mathcal{U}_{33}}-\frac{\mathcal{D}_{23}^\ast}{\mathcal{D}_{33}^\ast}\right)\epsilon_L^{23} & 1 
\end{pmatrix}}\,\nonumber
\end{align}
\end{widepageII}
assuming that both $\epsilon_L^{12}$ and $\epsilon_L^{23}$ are small.
The model therefore predicts that the product of the Cabibbo angle ($\sim\epsilon_L^{12}$) and the 2-3 quark mixing angle ($\sim\epsilon_L^{23}$) is of order the 1-3 mixing angle, a relation that is approximately correct. Indeed, to leading order the CKM matrix in our model can be matched onto the Wolfenstein parametrization~\cite{PhysRevLett.51.1945}. The Cabibbo angle $\lambda \approx 0.23$ is given by the combination
\be \label{eq:Cabibbo}
\lambda \approx |V_{us}| = \epsilon_L^{12} \left|\left(\frac{\widehat{\mathcal{D}}_{21}}{\widehat{\mathcal{D}}_{11}}-\frac{\widehat{\mathcal{U}}_{21}^\ast}{\widehat{\mathcal{U}}_{11}^\ast}\right)\right|\, .
\ee
The CKM matrix element $V_{cb}$, which is empirically observed to be of order $\lambda^2$, is given by 
\begin{align} \label{eq:Vcb}
V_{cb} = \epsilon_L^{23} \left(\frac{\mathcal{D}_{23}}{\mathcal{D}_{33}}-\frac{\mathcal{U}_{23}^{\ast}}{\mathcal{U}_{33}^{\ast}}\right)\, .
\end{align}
The scaling of our small parameters with the Cabibbo angle is then
\be 
\epsilon_L^{12} \sim \lambda, \qquad \epsilon_L^{23} \sim \lambda^2\, ,
\ee
assuming the fundamental parameters are chosen so that the prefactors appearing inside brackets in Eqs.~(\ref{eq:Cabibbo}) and (\ref{eq:Vcb}) are $\cO(1)$. We hereon set $\epsilon_L^{12} = \lambda$ and $\epsilon_L^{23} = \lambda^2$ precisely, without losing any freedom in our ability to fit all the mass and mixing parameters.

Plugging these back into the formulae for the mass eigenvalues, we predict the rough relations between the families,
\begin{align} \label{eq:m-ratios}
\frac{m_2}{m_3} &\sim \lambda^2 \epsilon_R^{23} \, , \\
\frac{m_1}{m_2} &\sim \lambda \epsilon_R^{12} \, .
\end{align}
Thus, having fit the hierarchies in the measured CKM angles via $\epsilon_L^{12}$ and $\epsilon_L^{23}$, the remaining expansion parameters $\epsilon_R^{12}$ and $\epsilon_R^{23}$ can be fit to the general trend in the mass ratios between generations, as observed across up, down, and charged lepton. (The residual differences between these species will be absorbed by the EFT coefficients.) For example, by crudely comparing (\ref{eq:m-ratios}) with the geometric means of the fermion masses in each family, one favours 
\be \label{eq:Rvev_scaling}
\epsilon_R^{12}\sim \lambda^2, \qquad \epsilon_R^{23}\sim \lambda\, .
\ee
In other words, the right-handed mixing angles can exhibit the `opposite' scaling to the left-handed mixing angles, so that the mass hierarchies $m_1/m_2\sim \lambda^3$ and $m_2/m_3 \sim \lambda^3$ are roughly equal.

Note that if we sub these scaling relations back into the hierarchical form (\ref{eq:Mmatrix}) predicted for the Yukawa matrices, we can express the hierarchical structure (up to the order-1 coefficients) in terms of the Cabibbo angle $\lambda$, finding:
\begin{align} \label{eq:M_lambda}
m
\sim \frac{v}{\sqrt{2}}
\begin{pmatrix}
\lambda^6 & \lambda^4 & \lambda^3 \\
\lambda^4 & \lambda^3 & \lambda^2 \\
\lambda^3 & \lambda & 1\\
\end{pmatrix} \, ,
\end{align}
for up-type quarks, down-type quarks, and charged leptons. 

\subsubsection*{Relations in the CKM matrix, and the Jarlskog invariant}

Our formula (\ref{eq:I}) for the CKM matrix does not parametrize a completely general unitary matrix. But happily, as mentioned above, our model can be matched onto the Wolfenstein parametrization. For example, both our Eq.~(\ref{eq:I}) and the Wolfenstein parametrization of the CKM matrix satisfy  the following relations:
\begin{align}
V_{cd}&=-V_{us}^\ast,\label{eq:Vus} \\ 
V_{ts}&= -V_{cb}^\ast, \label{eq:Vts}\\
V_{ud}&=V_{cs}^\ast, \label{eq:udcsEq}\\
V_{us}V_{cd}&=(V_{ud}+V_{cs}-2),\label{eq:us_relation}\\
V_{td}+V_{ub}^\ast &= (V_{us} V_{cb})^\ast\, , \label{eq:J-relation}
\end{align}
which are not implied by unitarity alone.
The first three relations imply $|V_{cd}|=|V_{us}|$, $|V_{ts}|=|V_{cb}|$, and $|V_{ud}|=|V_{cs}|$, all of which agree well with experiments.
Moreover, relations  (\ref{eq:udcsEq}) and (\ref{eq:us_relation}) together imply 
\begin{align}\label{eq:VudintermsofVus}
|V_{ud}|=|V_{cs}| = 1-\frac{1}{2}|V_{us}|^2+\ldots,
\end{align}
where `$+\ldots$' indicates terms suppressed by factors of $\lambda$. This too agrees well with observation.

Using a Gr\"obner basis analysis, the relation (\ref{eq:J-relation}) can be combined with the definition of the Jarlskog invariant, $J=\mathrm{Im}(V_{us} V_{cb} V_{ub}^\ast V_{cs}^\ast)$, to give
\begin{align}\label{eq:J-realfull}
4J^2= 2|V_{us}V_{cb}|^2( |V_{ub}|^2+|V_{td}|^2) + 2|V_{ub}V_{td}|^2 - |V_{td}|^4- |V_{ub}|^4 - |V_{us}V_{cb}|^4\, ,
\end{align}
at leading order in $\lambda$.
Note that this formula for $J$, which to the authors' knowledge is new,
indeed vanishes if all entries of $\CKM$ are real.
The Jarlskog invariant is a function of the $CP$-violating phase usually denoted by $\delta_{13}$ (in the standard parametrization of the CKM matrix).
Plugging into this relation the central observed values of the CKM angles $\theta_{12}$, $\theta_{13}$ and $\theta_{23}$ and solving for $\delta_{13}$ gives 
\begin{align}
\delta_{13}\approx 1.25~\mathrm{(radians)}\, ,
\end{align}
which agrees well with the experimental value.
Relation (\ref{eq:J-relation}) also implies the following inequalities,
\begin{align}\label{eq:inequality}
|V_{ub}|^2+|V_{cb}V_{us}|^2-2|V_{ub}V_{cb}V_{us}|<|V_{td}|^2<|V_{ub}|^2+|V_{cb}V_{us}|^2+2|V_{ub}V_{cb}V_{us}|\, .
\end{align}
These are indeed satisfied by the central values of the observed CKM matrix. 

It is not hard to see that if we can fit $V_{us}$, $V_{cb}$ and $V_{ub}$ to any complex numbers, we can of course fit $|V_{us}|$, $|V_{cb}|$ and $|V_{ub}|$ freely, and we can fit $|V_{td}|$ to any value satisfying (\ref{eq:inequality}). This then determines the CP-phase through (\ref{eq:J-realfull}), the value of $|V_{ud}|=|V_{cs}|$ through (\ref{eq:VudintermsofVus}), and the values of $|V_{cd}|=|V_{us}|$ and  $|V_{ts}|=|V_{cb}|$. Thus, by fitting 
$|V_{us}|$, $|V_{cb}|$, $|V_{ub}|$ and  $|V_{td}|$ to their central experimental values, we will reproduce the remaining $\CKM$ observables in close agreement with their experimental values.

\subsubsection*{Parameter space of the model}

Although we do not explicitly include every possible contribution in the scalar potential (\ref{eq:V}), our EFT nevertheless has enough freedom to fit (in complex space) all quark and charged lepton masses as well as the three CKM elements $V_{us}$, $V_{cb}$ and $V_{ub}$, as follows. (And as just discussed, this in turn allows us to fit the whole CKM matrix to a very good approximation.)
\begin{itemize}
\item Firstly, $\{y_{t},y_b,y_{\tau}, V_{cb}\}$ have no dependence on couplings to $\Phi_R$, and can be fit using the Yukawa coefficients $\{y_{1},\overline{y}_{1},y_{15},\overline{y}_{15}\}$ for any values of $(\beta_L^1,\beta_L^{15})$ away from a small set of points. The reader can easily verify this, since $V_{cb}$ is simple enough to write explicitly in terms of fundamental parameters; from Eq.~(\ref{eq:Vcb}), we have
\begin{align}
V_{cb} =&\frac{\lambda^2}{2}\Big\{\frac{\beta_L^1}{y_b}(y_1v_1+\overline{y}_1 \overline{v}_1^\ast)+\frac{\beta_L^{15}}{y_b}(y_{15}v_{15}+\overline{y}_{15} \overline{v}_{15}^\ast)\nonumber \\-&\frac{\beta_L^1}{y_t^\ast}(\overline{y}_1^\ast v_1+y_{1}^\ast\overline{v}_{1}^\ast)-\frac{\beta_L^{15}}{y_t^\ast}(\overline{y}_{15}^\ast v_{15}+y_{15}^\ast\overline{v}_{15}^\ast)\Big\},
\end{align}
and $y_t$, $y_b$ and $y_\tau$ are given by the linear combinations (\ref{eq:3rd-fam-masses}).\footnote{Unsurprisingly, the $V_{cb}$ angle vanishes in the limit where both cubic couplings $\beta_L^1$ and $\beta_L^{15}$ vanish, since in this case the tree-level diagrams in Fig.~\ref{fig:dim5} are not generated at all.}
\item Secondly, the observables $\{y_c, y_{s}, y_\mu,  V_{us}, V_{ub}\}$ have a dependence on couplings in the scalar potential that contain at most one $\Phi_R$ field. While the analytic expressions for these observables are unwieldy, we find that four of them can be fit using the model parameters $\{\beta_R^1,\beta_{LR}^1, w_{23},\overline{w}_{23}\}$. There is a remaining linear combination of these observables which is in fact independent of any coupling to $\Phi_R$, and this can be fit using $\{\beta_{LL}^1,\beta_{LL}^{15}\}$. Note that both  $\beta_{LL}^1$ and $\beta_{LL}^{15}$, which are real couplings, are required to fit the complex observable.
\item \sloppy Finally, the remaining set of observables we need to fit are the first family fermion Yukawa couplings $\{ y_u,y_d,y_e\}$. These can be fit using the model parameters $\{ w_{12},\overline{w}_{12},\beta_{RR}^1\}$.
\item The remaining model parameters can then be chosen arbitrarily, except on certain lower-dimensional surfaces in parameter space on which certain observables can't be fit. 
\end{itemize}

\section{Conclusion} \label{sec:Conclusion}

In this paper we have introduced a fundamental gauge theory of flavour based on the gauge group $\FSS$. This gauge group, which is a family-enriched generalisation of the Pati--Salam symmetry, unifies electroweak and flavour symmetries in the UV. All three generations of SM+3$\nu_R$ fermions are packaged into two fields in the UV. Additional scalar fields are required to break this much-enlarged gauge symmetry down to the SM, and we have chosen an almost-minimal set $\{S_L, S_R, \Phi_L, \Phi_R\}$ of scalars, as listed in Table~\ref{tab:Scalars}, to do the job. The SM Higgs, as in the one-family Pati--Salam model, is embedded in a pair of bifundamentals of the extended electroweak symmetry, which transform in the singlet and adjoint of the $SU(4)$. The vevs of these Higgs fields now automatically carry a direction in flavour space, which we suppose is aligned with the heavy third family fermions.

As well as explaining the origin of three generations in terms of broken gauge symmetries that extend the electroweak force, we also find that such a model can naturally account for the very particular hierarchical structure observed in quark and charged lepton masses and quark mixings. Only the third family fermions have renormalisable Yukawa couplings, and the mechanism for generating the light Yukawas is simple. First, two of the scalars $S_L$ and $S_R$ acquire vevs that partially break $\FSS$ down to a family non-universal intermediate gauge symmetry -- see Fig.~\ref{fig:breaking}. Next, the extra components of the Higgs fields are integrated out at a heavy scale $M_H$, before the remaining scalars $\Phi_{L,R}$ condense. At this point, higher-dimension Yukawa-like operators are generated for all the light fermions, due to scalar interactions between the Higgses and $\Phi_{L,R}$. We compute all these EFT coefficients explicitly. Finally, once the remaining dynamical BSM scalars $\Phi_L$ and $\Phi_R$ acquire their vevs, breaking the gauge group down to the SM, the EFT Yukawa-like operators match onto the dimension-4 Yukawa operators of the SM. As long as the vevs of $\Phi_{L,R}$ are an order of magnitude or so lighter than $M_H$, all the observed hierarchies in the Yukawa matrices are generated for $\cO(1)$ fundamental couplings, and there is enough parametric freedom in the EFT coefficients to fit all the quark and charged lepton masses, as well as the complete CKM matrix (although not a generic unitary matrix).

\subsection*{Further work}

Concerning the possible phenomenological consequences of electroweak flavour unification, 
we have only begun to scratch the surface in this paper. We envisage a number of extensions of this work.
\begin{enumerate}
\item We have implicitly assumed that the scales associated with breaking the extra gauge symmetries are all suitably high. In this high-scale unification scenario, the low energy physics is essentially described by the SM plus three extra Higgs doublets with natural masses $\lesssim 1$ TeV, as in other Pati--Salam based models. 
More interestingly, since there is no danger of excessive proton decay in the model we have set out, the scale of electroweak flavour unification could be rather low (much lower than the traditional GUT scale). In this case, the complicated spectrum of extra gauge bosons, most of which mediate flavour-dependent forces (with the exception of the $U_1$ leptoquark), would have a rich low-energy phenomenology. By measuring up against the current flavour data, one could find the lowest possible scale of electroweak flavour unification, and the BSM effects that one would expect to see first.
\item Related to this, it is intriguing that hints of flavour-dependent forces are already being seen in the decays of $B$-mesons, predominantly by the LHCb collaboration. These `$B$-anomalies' are now observed with a high statistical significance in neutral current $b\to s \ell\ell$ transitions, and there remain hints of anomalies in $b \to c \tau \nu$ charged currents. An obvious question to ask is: is there a viable parameter space in which any of the heavy gauge bosons predicted by electroweak flavour unification can simultaneously explain the $B$-anomalies, while remaining consistent with all flavour bounds and related constraints? 
\item For the $b\to s \ell\ell$ anomalies at least, there is indeed an obvious candidate already present in our model, which is the $\ZP$ boson in the $SU(2)$ triplet that arises from $\phi_L^{23}$ acquiring its vev. This $\ZP$ couples only to left-handed quarks and leptons in the 2nd and 3rd families. Moreover, since its vev is suppressed with respect to the vev of $\phi_L^{12}$ by a factor of the Cabibbo angle, in order to explain the hierarchy in quark mixing angles $|V_{cb}| \sim \lambda |V_{us}|$, this $\ZP$ boson can be the lightest BSM gauge boson in the spectrum. We save a detailed phenomenological study of this possibility for future work.
\item We have not explained neutrino masses at all in this work. Simultaneously explaining the anarchic neutrino mixing angles using some combined seesaw mechanism, while preserving the explanation of hierarchies in the quark sector, poses some challenges. Such an explanation will warrant an extension of the field content of the model beyond the very minimal setup we have used here.
\item There are possible cosmological consequences of electroweak flavour unification that we have not explored. Firstly, the symmetry breaking $\GUV \to \GSM$ ought to give rise to monopoles because the homotopy group $\pi_2(\GUV/\GSM) \cong \Z$ is non-vanishing.\footnote{This can be easily seen from the long exact sequence in homotopy groups for the fibration $\GSM \to \GUV \to X$, where $X \cong \GUV/\GSM$. The portion
$$\dots \to \pi_2(\GUV) \to \pi_2(X) \to \pi_1(\GSM) \to \pi_1(\GUV) \to \dots $$
reads
$$\dots \to 0 \to \pi_2(X) \to \Z \to 0 \to \dots,$$
where we have used the fact that both $\pi_1$ and $\pi_2$ vanish for any compact semi-simple Lie group, while $\pi_1(\GSM)\cong \Z$ because of the hypercharge factor.
We read off that the group we want is $\pi_2(X) \cong \Z$.
}
This is the case for {\em any} UV model with semi-simple gauge group. One way to avoid an over-density of monopoles is for the scale $\Lambda_R$ of $SU(4)\times Sp(6)_R \to SU(3) \times Sp(4)_R \times U(1)_R$ breaking to be higher than the scale of inflation, meaning that the monopole density would be diluted during inflation.
Since there is no monopole production below the scale $\Lambda_R$, no more monopoles are produced after inflation ends.
\item Secondly, it was shown in Ref.~\cite{Greljo:2019xan} that the generation of flavour hierarchies via a succession of symmetry breaking steps can lead to a multi-peaked stochastic gravitational wave (GW) signal, that could in principle be detected by future interferometer experiments such as LISA, the Einstein Telescope (ET), and Cosmic Explorer (CE). If the analogous symmetry breaking steps in our model can proceed via first order phase transitions, there could be detectable GW signatures in our model also -- even in the case of very high scale symmetry breaking. Since in this scenario
there is little hope of directly producing the heavy gauge bosons at present or future colliders,
the detection of GWs could provide a complementary probe of electroweak flavour unification.
\end{enumerate}

\section*{Acknowledgments}

We are grateful to Benjamin Allanach and Ben Gripaios for discussions in the early stages of this project. JD thanks Darius Faroughy, Admir Greljo, Gino Isidori, Nakarin Lohitsiri, Ben Stefanek, and Anders Eller Thomsen for discussions. JTS thanks Matthias Linster and Robert Ziegler for correspondence and Maximilian Ruhdorfer for discussions. We further thank Admir Greljo and Gino Isidori for their careful reading of a draft of this manuscript.

JD has received funding from the
European Research Council (ERC) under the European Union’s Horizon 2020 research and innovation programme under grant agreement 833280 (FLAY), and by the Swiss National Science Foundation (SNF) under contract 200020-204428. JTS was supported in part by the U.S. National Science Foundation (NSF) grant PHY-2014071. Part of this work was carried out at the University of Cambridge, under the support of the STFC consolidated grants ST/P000681/1, ST/T000694/1 and ST/S505316/1.

\appendix

\section{Anomaly cancellation} \label{app:global}

In this Appendix we show that the $\GUV=\FSS$ gauge model is anomaly-free.

\paragraph*{No local anomalies.}

One must first consider local (perturbative) gauge anomalies.
\sloppy Since $Sp(6)$ has only real and pseudoreal representations there are no associated local anomalies, and it remains only to check that the $SU(4)$ factor is free of perturbative anomalies. This follows automatically, since there are equal numbers of left-handed and right-handed Weyls transforming in the anomalous fundamental representation of $SU(4)$.

\paragraph*{No global anomalies.}

\sloppy Having first checked that there are no local anomalies, one should next ask whether there are more subtle global anomalies in our UV gauge model. To detect global anomalies in a 4d gauge theory with gauge group $\GUV$, and with fermions defined using a standard spin structure, one should compute the torsion part of the bordism group $\Omega_5^\mathrm{Spin}(B\GUV)$.
This bordism group can be easily computed using the Atiyah--Hirzebruch spectral sequence (AHSS). We refer the curious reader to Refs.~\cite{Garcia-Etxebarria:2018ajm,Davighi:2019rcd} for similar examples of the AHSS in the context of BSM physics, and for an introduction to this method.\footnote{Ref.~\cite{Wan:2020ynf} uses the Adams spectral sequence to compute anomalies relevant to BSM physics.} 

The calculation of  $\Omega_5^\mathrm{Spin}(B\GUV)$ goes as follows.
Recall that, when $BG$ sits inside a fibration
$F \to BG \to B$,
the second page of the (homological version of the) AHSS is given by
$E^2_{p,q} = H_p\big(B; \Omega_q^\mathrm{Spin}(F)\big)$.
It here suffices to consider the trivial fibration $\mathrm{pt} \to B\GUV \to B\GUV$ of $B\GUV$ over itself, where $\mathrm{pt}$ denotes a point, 
meaning the input to the spectral sequence is
\be
E^2_{p,q} = H_p\left(B\GUV; \Omega_q^\mathrm{Spin}(\mathrm{pt})\right).
\ee
Recall that the first few bordism groups of a point are~\cite{anderson1966spin}
\begin{equation}
\label{eq:bordism-table}
\def\arraystretch{1.5}
\arraycolsep=4pt
\begin{array}{c|cccccccc}
n & 0 & 1 & 2 & 3 & 4 & 5 & 6 \\
\hline
  \Omega^{\text{Spin}}_n(\text{pt}) & \Z & \Z_2 & \Z_2 & 0 & \Z & 0 & 0 \\
\end{array} \, ,
\end{equation}
and so we need the first few homology groups of $B\GUV$ valued in $\Z$ and $\Z_2$, in order to populate $E^2_{p,q}$.

\begin{figure}[h]
\centering
  \includegraphics[width=0.65\textwidth]{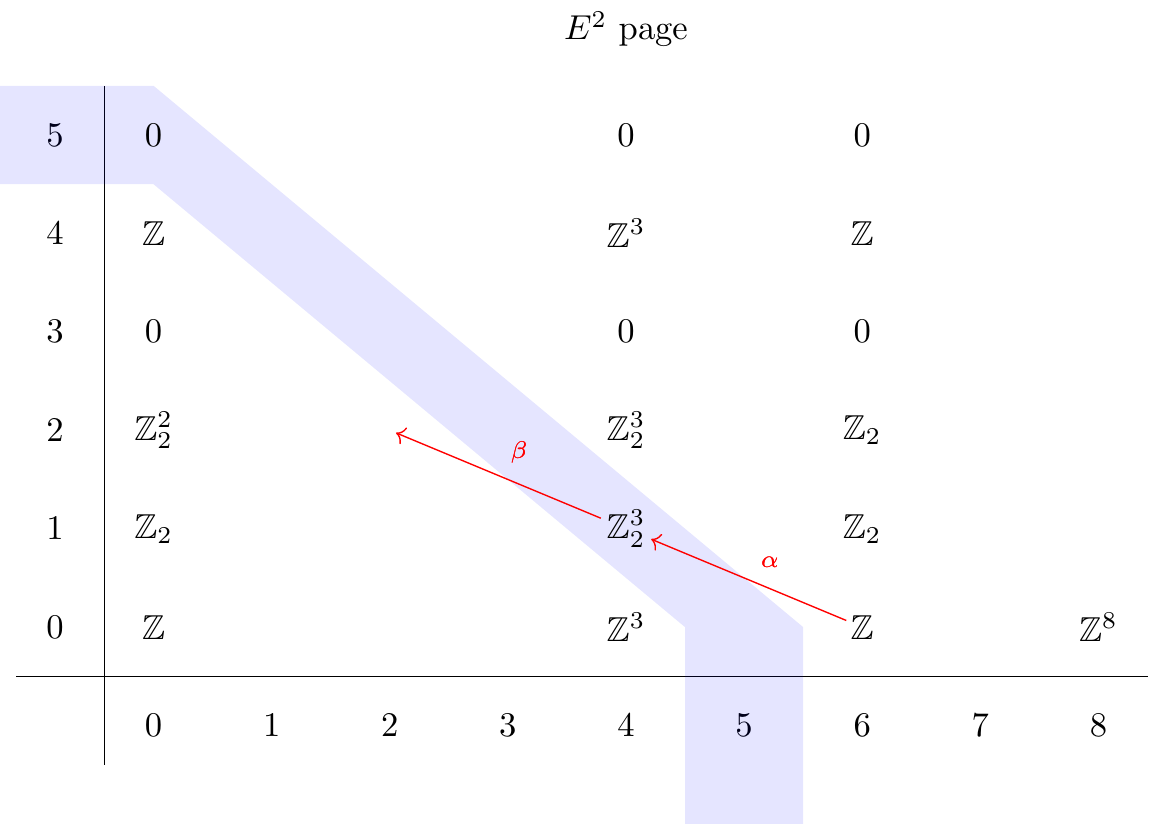}
   \label{fig:AHSSSU3}
\caption{The second page of the AHSS for $\Omega^{\mathrm{Spin}}_5\left(B\GUV\right)$. } \label{fig:AHSS}
\end{figure}

Such groups are easy to find. Firstly, the cohomology ring of $B\GUV$ is
\be
H^\bullet (B\GUV; \Z) \cong \Z \left[c_2, c_3, c_4, p_1^L, p_2^L, p_3^L, p_1^R, p_2^R, p_3^R \right]\, ,
\ee
where $c_i$ are the non-zero Chern classes of the $SU(4)$ factor, and $p^{L,R}_i$ are the non-zero Pontryagin classes of the $Sp(6)_{L,R}$ factors. Therefore, the low-dimension cohomology groups are $H^\bullet(B\GUV; \Z) \cong \{\Z,0,0,0,\Z^3,0,\Z,0,\Z^8,\dots \}$, and these coincide with the low-dimension homology groups by universal coefficients (since all the odd-degree cohomology groups vanish, and there is no torsion). Using this information, we write down the second page of the AHSS in Fig.~\ref{fig:AHSS}.

Already on the second page, there is only a single element on the $p+q=5$ diagonal that could contribute to $\Omega_5$, which is $E_{4,1}^2$. The map out, $\beta:E_{4,1}^2 \to E_{2,2}^2$ is the zero map so $\mathrm{ker}~ \beta \cong (\Z_2)^3$, and it remains only to compute the map in, $\alpha:E_{6,0}^2\to E_{4,1}^2$. This map is the composition of the dual of a Steenrod square and reduction modulo 2, 
\be
\alpha:\Z \xrightarrow{\text{mod 2}} \Z_2 \xrightarrow{\widetilde{\mathrm{Sq}^2}} (\Z_2)^3\, ,
\ee
where the Steenrod square that we need is simply
\begin{align}
\mathrm{Sq}^2:H^4(B\GUV; \Z_2) &\to H^6(B\GUV; \Z_2): \\
c_2 &\mapsto c_3, \quad p_1^L, \, p_1^R \mapsto 0. 
\end{align}
Its dual therefore sends $\widetilde{c_3}$ to $\widetilde{c_2}$ (where the notation denotes the dual to the corresponding elements in mod 2 cohomology), and so $\mathrm{Im}~ \alpha \cong \Z_2$, generated by  $\widetilde{c_2}$. Thus, taking the homology with respect to the differentials $\alpha$ and $\beta$, we turn to the next page:
\be
E^3_{4,1} \cong \frac{(\Z_2)^3}{\Z_2} \cong (\Z_2)^2 \, .
\ee
Continuing to turn pages, there are no further relevant differentials until the fourth page, but here the differential is a map $\gamma:E_{4,1}^4 \to E_{0,4}^4:(\Z_2)^2 \to \Z$ which must be the zero map. Hence, $E_{4,1}^\infty \cong (\Z_2)^2$. Since there are no other entries on the $p+q=5$ diagonal, we read off the fifth bordism group to be
\be
\Omega_5^{\mathrm{Spin}}\left[B\left(\FSS \right) \right] \cong \Z_2 \times \Z_2\, ,
\ee
which is pure torsion.

From the computation of the AHSS above, we see that these $\Z_2$ factors derive from the combination of each $Sp(6)_{L,R}$ Pontryagin class $p_1^{L,R}$, together with $\Omega^\mathrm{Spin}_1(\mathrm{pt}) \cong \Z_2$. 
Thus, in a 4d theory with gauge group $\FSS$, there is a pair of possible global anomalies, one associated with each $Sp(6)$ factor. These anomalies, which afflict any $Sp(2r)$ gauge theory (as was observed in Witten's first paper concerning global anomalies~\cite{Witten:1982fp}) are avatars of the more famous $SU(2)\cong Sp(2)$ anomaly. For one way to see this anomaly (see {\em e.g.}~\cite{Wang:2018qoy,Davighi:2020uab}), consider a single $Sp(6)$ fundamental fermion on a spacetime $M \cong S^4$, in the presence of an odd instanton, {\em i.e.} an $Sp(6)$-principal bundle over $M$ with $p_1([M])\in (2\Z+1)$. The fermion partition function flips sign under the gauge transformation by $-{\bf 1}\in Sp(6)$,\footnote{
This can be seen by constructing a 
5d mapping torus $X_5 \cong M \times S^1$, where the spin structure has periodic boundary conditions around the circle that implement a $(-1)^F$ twist, or equivalently a constant gauge transformation by $-{\bf 1}\in Sp(6)$, and evaluating the Atiyah--Patodi--Singer $\eta$-invariant
on this mapping torus (see~\cite{Witten:2019bou} and references therein). Needless to say, this 5d mapping torus is a representative of the non-trivial class in $\Omega_5^\mathrm{Spin}(BSp(6))$, and evaluating the $\eta$-invariant, which is here a cobordism invariant, gives a phase of $-1$ for our choice of Dirac operator.
} meaning that $Sp(6)$ is anomalous.

In general, each $Sp(6)_{L,R}$ anomaly is generated by odd numbers of fermions transforming in the fundamental $\mathbf{6}$ representation (amongst others\footnote{A representation $\mathbf{R}$ will contribute to the anomaly if the Dynkin index $T_{\bf{R}}$ of the representation is an odd number, where this Dynkin index is defined via $\Tr (t^a_{\bf{R}} t^b_{\bf{R}})=\frac{1}{2} T_{\bf{R}} \delta^{ab}$, with $\{t^a_{\bf{R}}\}$ being a a basis of $Sp(6)$ generators in the representation ${\bf R}$.
}) of either $Sp(6)$. Since our electroweak flavour unification model features an even number (four) of Weyl fermions transforming in the fundamental representation of each of $Sp(6)_L$ and $Sp(6)_R$, both global anomalies cancel, exactly as for the ordinary Pati--Salam model.
We conclude that our $\FSS$ gauge model is completely anomaly-free on any suitable spin 4-manifold equipped with gauge bundle.

\bibliography{refs_466}
\bibliographystyle{JHEP}
\end{document}